\documentclass[%
reprint,
 superscriptaddress,
nofootinbib,
 amsmath,amssymb,
pra,
floatfix,
]{revtex4-2}

\usepackage[english]{babel}
\usepackage{mathtools}
\usepackage{float} 
\usepackage{array}
\usepackage{tikz}
\usetikzlibrary{arrows.meta}

\usepackage{lipsum} 
\usepackage{graphicx}%
\usepackage{dcolumn}%
\usepackage{bm}%
\usepackage[hidelinks,colorlinks=true,linkcolor=blue,citecolor=blue]{hyperref}

\usepackage{dsfont}
\usepackage{array}
\newcolumntype{P}[1]{>{\centering\arraybackslash}p{#1}}
\usepackage{adjustbox}

\usepackage{braket}

\usepackage{amsmath}
\usepackage{booktabs}
\usepackage{tabularx}
\usepackage{array}
\usepackage{todonotes}
\usepackage[noabbrev]{cleveref}

\makeatletter
\let\originalFMmark\frontmatter@thefootnote
\renewcommand{\frontmatter@thefootnote}{%
  \ifnum\value{\@mpfn}=2\relax
    \ensuremath{\circ}
  \else
    \originalFMmark
  \fi}
\makeatother

\makeatletter
\newcommand\nextcolumnfloat{\global\@colnum\z@}
\makeatother

\begin{document}

\preprint{APS/123-QED}

\title{Millimetre-scale experiments can test the quantum nature of spacetime}
\author{Lorenzo Braccini$^\dagger$}
\email{lorenzo.braccini@gmail.com}
\affiliation{Department of Physics and Astronomy, University College London, Gower Street, WC1E 6BT London, United Kingdom}

\author{Emanuele Panella$^\dagger$}
\email{emanuele.panella@uniroma2.it}
\affiliation{Dipartimento di Fisica, Università di Roma Tor Vergata and Sezione INFN Roma 2, Via della Ricerca Scientifica 1, 00133,
Roma, Italy}

\author{Alessio Serafini}
\affiliation{Department of Physics and Astronomy, University College London, Gower Street, WC1E 6BT London, United Kingdom}

\author{Sougato Bose}
\affiliation{Department of Physics and Astronomy, University College London, Gower Street, WC1E 6BT London, United Kingdom}

\author{Jonathan Oppenheim}
\affiliation{Department of Physics and Astronomy, University College London, Gower Street, WC1E 6BT London, United Kingdom}

\begingroup
\renewcommand{\thefootnote}{$\dagger$}
\footnotetext{These authors contributed equally to this work.}
\endgroup

\begin{abstract} 
Observing gravitationally mediated entanglement (GME) is widely regarded as a witness of the quantum nature of gravity. Yet theories in which gravity is classical can contain nonlocal interactions, and the exact general conditions under which these could produce the same signal have not been established.  We derive them for all continuous, Markovian theories that couple classical gravity to quantum matter. Such theories must be constrained by a minimal noise floor due to a trade-off between classical randomness and quantum decoherence, set by the strength of the gravitational force. We prove that this trade-off forbids entanglement between masses that can be separated by a plane, unless the density of the two masses is smeared at the same scale as their separation. This weakens the Newtonian force between them by an order of magnitude, meaning that force measurements at the GME separation, accurate only to tens of percent, can exclude this mechanism. The same experiments also constrain how far smearing, or running of Newton's constant, can lower the predicted noise floor. Existing short-range experiments already limit a change in the strength of gravity  to below the millimetre scale, which implies an acceleration-noise floor about five orders of magnitude beneath LISA Pathfinder's sensitivity. Force, noise, and decoherence experiments are therefore nearer-term milestones toward GME, which would exclude smearing-induced entangling mechanisms and, importantly, rule out a large class of theories in which spacetime remains classical.
\end{abstract}

\maketitle

\section{Introduction}

Recently, there has been a resurgence of interest in experimentally testing the fundamental nature of the gravitational field.
Unlike the other forces, gravity in general relativity is the curvature of spacetime itself, and spacetime is not simply one more quantum field, but the arena in which quantum states are defined and evolve. Whether it must be quantised has therefore long been regarded as an open question, and ultimately one for experiment~\cite{rosenfeld1963quantization,unruh1984steps,carlip2008quantum,dyson_is_2013,oppenheim2023it}.
Experiments have been proposed with the common aim of distinguishing quantum from classical gravity at low energies~\cite{bose_matter_2016,bose_spin_2017,marletto_gravitationally-induced_2017,carney2019tabletop,howl_non-gaussianity_2021,biswas_gravitational_2022,lami_testing_2024,tobar2024detecting,hanif2024testing,kryhin2025distinguishable,sharmila2026signatures,Oppenheim:2022xjr,saldanha2026repulsive,bose2026effective,das2026test}, thereby avoiding the formidable challenge of probing the Planck scale directly~\cite{dyson_is_2013}. A prominent class of proposals focuses on the detection of gravitationally mediated entanglement (GME) between two spatially superposed masses~\cite{bose_spin_2017,marletto_gravitationally-induced_2017}, relying on the fact that separable states cannot become entangled under Local Operations and Classical Communication (LOCC)~\cite{werner1989quantum,bennett_mixed_1996}. Under the assumption that a classical gravitational mediator can transmit only classical information through local interactions, the LOCC theorem implies that the generation of entanglement would witness non-classical features of gravity \cite{bose_spin_2017,marshman_locality_2020}.

Guaranteeing the right conditions for such entanglement detection  would be a remarkable experimental achievement. The interacting masses must be exceptionally coherent, highly delocalized, and sufficiently close to acquire significant GME. This demands outstanding isolation from environmental decoherence and screening from short-range electromagnetic and Casimir interactions~\cite{Romero_Isart_quantum_2011,toros_relative_2021,toros_creating_2021,pedernales_motional_2020,grossardt_2020_acceleration,bose_spin_2025,braccini2026massindependentgravitationallyinducedentanglement,bose2026white,van_de_kamp_quantum_2020,schut_relaxation_2023,schut_micronsize_2023,bulling_stability_2026}. A complete experiment must satisfy all of these requirements and must ultimately implement an entanglement witness (rather than merely detect an average force)~\cite{chevalier_witnessing_2020,bose_spin_2017}. Many physical systems have been proposed for the detection of GME~\cite{bose_spin_2017,weiss_large_2021,al_optomechanical_2018,matsumura_gravity-induced_2020,krisnanda2020observable,haine_searching_2021,carney_using_2021,agafonova2025onemill} and great experimental effort is currently undergoing to show quantum delocalization of large masses~\cite{rossi_levitodynamics_2026,Quantum_Kamba_2025,quantum_rossi_2025,kremer_2026_fast}.

Several computations in low-energy quantum gravity show that GME is a robust prediction of a quantum-natured gravitational interaction~\cite{belenchia_quantum_2018,christodoulou_possibility_2019,marshman_locality_2020,carney_newton_2022,chen_quantum_2022,bose_mechanism_2022,christodoulou_locally_2022,galley_no-go_2022,biswas_gravitational_2022,danielson_gravitationally_2022,bengyat_gravity_2023}, even in the presence of noise. However, in classical theories of gravity, the validity of the LOCC argument is less clear. It rests on the assumption that there always exists a representation of the dynamical map of the matter, conditional on the state of the gravitational field, which is subsystem local~\cite{marshman_locality_2020,oppenheim2023covariant} (this is motivated by assuming that the field interactions are local in spacetime, i.e. micro-causality~\cite{circuit_dibiagio_2026,spalvieri2026localoperationsfieldmediated,Boulle_2026_subsystem}). For classical-quantum theories with local noise and local coupling to matter, such a representation exists and no entanglement is generated~\cite{oppenheim2023covariant}. However, this is not generically the case -- classicality of the gravitational field poses no such restriction~\cite{Tilloy:2015zya}. Furthermore, one never has access to the  conditional state, since the Newtonian potential itself is only accessible indirectly through the matter degrees of freedom, which serve as both the source and the sensor of gravity. After averaging over the state of the gravitational field, the quantum masses do not generically evolve under a map which is subsystem-factorizable and, hence, explicitly not entangling. The Newtonian limit of generic classical--quantum (CQ) theories~\cite{layton2023weak} features the nonlocal $1/\hat{r}$ quantum two-body interaction term, accompanied by a spatially correlated fundamental noise (noiseless coherent interaction cannot arise classically while assuming a mediator~\cite{kafri2014classical,bose_mechanism_2022,galley_no-go_2022}).

Explicit calculations have indeed shown that entanglement can arise in a popular theory of Newtonian classical gravity~\cite{trillo_diosi_2025,entanglement_angeli_2025}, the Di\'osi--Penrose (DP) model~\cite{Diosi:1986nu,Tilloy:2015zya,Penrose:1996cv}. In this context, recent analyses have discussed experimental constraints on the specific parameters that would yield entanglement under the DP model with Gaussian smearing~\cite{aspemeyer_quantum_2026}, as well as the role of the fundamental noise induced by the classicality of the field when the latter is fundamentally spacetime local~\cite{minimal_fabiano_2026}. These considerations allow one to exclude a \emph{restricted} subset of these theories by entanglement detection. However, a general statement is still lacking, and a key question remains: \emph{can a consistent classical theory generate GME in the proposed experimental setups, while being otherwise empirically indistinguishable from quantum gravity?} This is a particularly timely question, since on its answer relies the significance of the great efforts devoted to detecting GME -- and the counterexample of the DP model casts doubt on the discriminating power of such a challenging experiment. Answering it requires a criterion that remains valid beyond particular choices of gravitational noise and matter regularization.

It is important to clarify now what we mean by “classical” gravity. For the gravitational field to have that connotation, a minimal property to require is that it has an explicit state space, where allowed states are classical field configurations, or probability mixtures of them. Quantum matter, on the other hand, still needs to have a Hilbert space description and allow for linear superposition of states. The dynamics, then, should preserve the notion of probabilities, the classical-quantum split and the statistical interpretation of the density matrix (which requires the dynamics to be linear in the classical-quantum state).  This set of conditions that restricts the allowed models to those that can be described within the so-called CQ framework~\cite{oppenheim2021postquantum,oppenheim2022the-two-classes}, whose dynamics needs to be irreversible~\cite{Oppenheim:2022xjr,Galley:2023byb}. To the best of our knowledge, other mechanisms for the generation of GME with “classical” gravity in the literature (examples are~\cite{classical_aziz_2025,martin-martinez_what_2022}, of which the former was shown to be incorrect~\cite{marletto2025classical,diosi2025no,schneider2026demonstration,gundhi2026can}) fail to satisfy at least some of these minimal assumptions -- and so they are not well-defined classical models. “Semiclassical gravity”, a model which sacrifices the linearity property in quantum mechanics, has been shown not to entangle in its Newtonian limit~\cite{Struyve:2016obi,Fedida:2026xlp}.

Here we establish a condition under which theories in the broad class of continuous, Markovian classical–quantum models of Newtonian gravity cannot generate GME. Sec.~\ref{sec:master} introduces the effective dynamics on the masses, motivated by such classical theories, as well as by quantum gravity considerations. In Sec.~\ref{sec:entangling_condition}, we prove that the fundamental noise floor of classical gravity is a sufficient condition to prevent GME for masses separable by a plane, in the absence of smearing, irrespective of the spatial locality of such noise kernel. In Sec.~\ref{sec:CQ}, we include the effect of smearing, showing that these theories can entangle only if Newton's law breaks down at the GME scale; reducing the gravitational attraction between the masses by at least 84\%, independently of the choice of smearing function and noise kernel. This provides a direct route to exclude such a mechanism to produce gravitational entanglement in a classical model, through independent short-range force measurements (Sec.~\ref{subsec:r0_exp}). Applying our framework to existing precision measurements, we upper bound the smearing parameter at the millimetre scale. We further connect these force constraints to the minimum noise induced by classical–quantum gravity, which, in principle, can be lowered arbitrarily by increasing the smearing scale (Sec.~\ref{subsec:tradeoff}). The upper bound we derive on such non-local scale implies that LISA Pathfinder's differential-acceleration noise sensitivity lies approximately five orders of magnitude in amplitude above the corresponding theoretical minimal floor.  Observation of a smaller noise would provide the exciting opportunity to exclude any continuous and Markovian classical theory of gravitation. As discussed in  Sec.~\ref{subsec:milestone}, this is a concrete milestone on the way towards GME since, as we show, it is a necessary condition (although not sufficient)  for the noise affecting the masses to lie below this minimal floor for the dynamics to be entangling. Our results thus have two profound implications -- they bring the experimental investigation of classical gravity within closer reach, while establishing benchmarks needed to make future observations of GME a compelling signature of the quantum nature of the gravitational field.

\textit{Conventions:} Unless stated otherwise, in the main body we work in units of $\hbar=1$.

\section{Newtonian Gravity, \\ Noise \& Smearing }
\label{sec:master}

Almost all practical setups considered for detecting GME involve two non-relativistic matter distributions, with matter-density profiles $m_i(x)$ and total masses $M_i$ for $i=1,2$, which we shall also refer to as the `nanoparticles' in what follows. Relativistic effects are expected to be completely negligible in this regime; hence, we adopt a nonrelativistic description of the gravitational interaction between the masses, in which the gravitational field couples to the total mass-density operator $\hat{m}(x)=\sum_i \hat{m}_i(x)$. In practice, the only accessible quantities in a laboratory are the matter distributions themselves, which are both the sensor and the source of gravity -- the Newtonian gravitational field is not accessible directly, and is therefore integrated out in our description. Throughout, we take the dynamics to respect Galilean invariance. 

Hence, our starting point is the following master equation for the matter distribution 
\begin{equation}
\label{eq:master}
    \partial_t \hat{\varrho} = i[\hat{H}_{N},\hat{\varrho}] - \mathcal{D}_g(\hat{\varrho}) \ ,
\end{equation}
where $\hat{H}_{N}$ is the gravitational coherent interaction term in the Newtonian, non-relativistic limit
\begin{equation}
  \hat{H}_N =  \frac{1}{8\pi \kappa} \int {\rm d}^3 x \text{d}^3  y \frac{\hat m_g(x)\hat m_g(y)}{|x- y|} \;,
\end{equation}
where $\kappa = 1/(4 \pi G)$ and 
\begin{equation}
\label{eq:dissipator}
     \mathcal{D}(\hat{\varrho}) =  \frac{1}{2}\int {\rm d}^3 x \text{d}^3  y   D(x - y) \left[\hat{m}_g(x),[\hat{m}_g(y),\hat{\varrho}]\right] 
\end{equation}
is a Lindbladian term describing decoherence in the mass-density basis.
We have allowed the bare mass-density operator to be smeared by some positive, centred and normalised kernel
\begin{equation}
\label{eq:smearing}
    \hat{m}_g(x) = \hat{m} * g (x) = \int  \text{d}^3  y \ \hat{m} (y)  g (x - y) .
\end{equation} 
Eq.~(\ref{eq:master}) describes a large class of gravitational dynamics under arbitrary gravitational and matter noises with general mass-smearing. This type of dynamics arises frequently in both quantum and classical theories of gravitation in the Newtonian limit. 

On the one hand, environmental noise and a nonlocality scale need to be taken into account even in a purely quantum gravity theory. In practice, the value of the gravitational field at a point will fluctuate by some random amount $\delta\Phi(x)$, an effect that, although small, cannot be screened~\cite{toros_relative_2021,toros_loss_2020}. Treating those fluctuations as some unobservable bath, naturally leads to decoherence in the mass-density basis, in which $D$ would encode the correlations in the fluctuations of the gravitational field, alongside the many other sources of matter decoherence due to coupling to other baths (such as black-body environmental photons). Moreover, a variety of gravitational theories predict either a breakdown of the inverse-square law (e.g. ghost-free gravity~\cite{biswas2012towards,Frolov:2015bia}, or the “fat graviton” of finite size~\cite{sundrum2004fat} proposed at the sub-millimetre scale) or a fundamental UV regularisation of the mass-density operator (a typical effect in non-commutative geometry approaches~\cite{Smailagic:2003yb}). At the level of the Newtonian description, many UV regularisation of the Newtonian interaction have an equivalent description in terms of mass-density smearing; our analysis focuses on this case. However, analyses of GME typically neglect this nonlocality scale because quantum gravity models often associate it with the experimentally inaccessible Planck scale.

On the other hand, these modifications naturally appear in a wide range of classical gravity models, which we refer to as CQ gravity, where they play a much more prominent role. The set of CQ models includes all the theories of gravity in which the metric is fundamentally classical, under the assumptions that the generator of the dynamics is linear in the state, probability-preserving and (completely) positive. These theories feature fundamental stochasticity in the Newtonian potential, with covariance $W(x-y)$, and decoherence in the mass-density basis with kernel $\lambda(x-y)$. The two kernels need to obey the so-called decoherence--diffusion (DD) trade-off~\cite{Oppenheim:2022xjr}:
\begin{equation}
    W(k)\lambda(k) \geq 1 \ ,
\end{equation}
where we moved to $k$-space for simplicity\footnote{We use the following convention for the Fourier transform: $f(k) =\int d^3x\, e^{-ik\cdot x}f(x)$ and $f(x) =\int \frac{d^3k}{(2\pi)^3}\, e^{ik\cdot x}f(k)$.}.

Fixing the Newtonian interaction, and averaging over the states of the classical field, the reduced evolution of the matter degrees of freedom can be shown~\cite{Tilloy:2015zya} (we present our derivation in terms of path integrals in Appendix~\ref{app:dynamics}) to take the form of Equation~\ref{eq:master}, including the dissipation term. In fact, the DD trade-off imposes the extra constraint~\cite{tilloy2017principle}
\begin{equation}
\label{eq:dp_floor}
    D(k) = \lambda(k) + \frac{W(k)}{4\kappa^2 k^4} \geq \frac{1}{\kappa k^2} \equiv D_*(k)  \ ,
\end{equation}
providing a minimal noise floor -- the model with minimal decoherence kernel $D_*$ is precisely the Diosi-Penrose model. The dissipator is therefore non-negligible in fundamentally classical models of gravitation, contrary to the quantum case in which, with full control over the environment, one is fundamentally limited only by vacuum fluctuations of the gravitational field. Crucially, the classicality of gravity imposes no constraint on the locality of the noise kernels -- the minimal noise floor of Eq.~(\ref{eq:dp_floor}) is nonlocal with spatial correlations of the form $1/r$. Still, a model exists in which the fundamental noise kernels $W,\lambda\propto \delta^{(3)}(x-y)$ are local, although the combined effective decoherence on the matter is not:
\begin{equation}
\label{eq:ultra_local_kerenl}    D^{\mathrm{loc}}(k) = D_l+\frac{1}{4D_l \kappa^2 k^4} \geq D_*(k) \ \; \forall\; D_l> 0 \ .
\end{equation}
Here, $D_l$ can then  be tuned to minimise the noise for some chosen $k$.

It is not unreasonable that an intermediate length scale should exist also for CQ theories. In the absence of matter and of a cosmological constant, the action of the full postquantum theory contains no dimensionful parameter and the Planck length does not enter into the equation of motion: Newton's constant doesn't enter in the pure gravity theory only appearing through the coupling of spacetime to matter~\cite{grudka2024renormalisation}. If the dimensionless couplings of the gravitational action run, a new length scale can be generated dynamically, as happens in QCD, at a value to be fixed by experiments. As mentioned earlier, smearing the mass-density operator is equivalent to an effective gravitational coupling $G_{\rm eff}(k)=G|g(k)|^2$ between unsmeared mass densities, at the Newtonian level. This has an important consequence, since the diffusion–decoherence trade-off holds separately at each wavenumber $k$ and, at fixed $k$, the minimum noise in Eq.~(\ref{eq:dp_floor}) is proportional to $G_{\mathrm{eff}}(k)$ itself. Smearing therefore lowers the noise floor at precisely those wavenumbers where it suppresses the Newtonian interaction. Whenever the modification has an equivalent representation in terms of mass-density smearing, the gravitational force is nowhere stronger than Newtonian ($G(k)\leq G$), in contrast to, e.g., large extra dimensions~\cite{arkani--hamed1998hierarchy}. 

Another consequence of smearing is that it arbitrarily lowers the effective size of the stochastic forces on the quantum particles, reducing the large heating associated with these models~\cite{Ghirardi:1989bb}. A common choice of smearing function for the DP model is the Gaussian kernel
\begin{equation}
\label{eq:gaussian_smearing}
    g_{R_0}(x)=\frac{1}{(2\pi R_0^2)^{3/2}}e^{-|x|^2/(2R_0^2)} \ .
\end{equation}
To compare the effects of different regularisations of the mass-density operator, however, we extend the definition of $R_0$ beyond the Gaussian case. Following Ref.~\cite{Piccione:2025zfb}, it is useful to define $R_0$ as the standard deviation of the smearing function $g$ along any Cartesian direction
\begin{equation}
    R_0^2 \equiv \frac{1}{3}\int \text{d}^3x \ g(x)x^2 \ ,
\end{equation}
Great experimental efforts have been made to lower-bound the smearing scale with decay experiments, finding $R_0 \gtrsim 5\,\text{\AA}$ for the Gaussian kernel~\cite{XENON:2025iov}. Since this exceeds the typical interatomic spacing of solids, throughout this article we treat extended bodies as continuous and ignore their granularity~\cite{bassi_models_2013}. 

It is precisely the Gaussian smearing of the mass-density operator that has been linked, in the Diosi-Penrose model, to entanglement generation~\cite{trillo_diosi_2025,angeli_positivity_2025}. With a single modification, classical-quantum models seem to be able to escape two independent tests that had in principle the power to rule them out: detecting GME and probing the DD trade-off~\cite{Oppenheim:2022xjr}.

\section{The entanglement condition \label{sec:entangling_condition}}
In the case of rigid bodies (such as nanoparticles, torsion pendula, and mirrors), the matter distribution is fixed around the centres of mass described by the continuos variable quantum operators $\hat{X}_i$ and $\hat{P}_i$, with commutation relations $[\hat{X}_i, \hat{P}_j ] = i \delta_{ij} $,  ($i = 1,2$). These are the variables, accessible in the experimental setting via appropriate measurements, that may encode the non-classical correlations of GME. At the level of the matter distribution, this amounts to setting $\hat{m}_g(x)=M \sum_i\rho_g( x- \hat{X}_i)$, where, $\rho_g(x) = \rho * g (x)$ is the smeared density profile. By fixing $\rho$,  deformations during the experiment (such as non-Gaussianity of condensates~\cite{howl_non-gaussianity_2021}) are neglected. For now, we consider identical, bare, normalized matter-density distributions of total mass $M$ -- for instance, we approximate nanoparticles as identical spheres. This choice is for ease of presentation, and can be relaxed (see Appendix~\ref{app:unequal_masses}). 

Each particle is prepared in a spatial superposition of its centre of mass, with delocalisation scale $\Delta x$, collinear with the direction of the separation (which we align with the $z$-axis). The initial state is taken to be uncorrelated and hence described by the quantum state 
\begin{equation}
    \ket{++} =\frac{1}{2} \left(\ket{\Delta x/2}_1+\ket{-\Delta x/2}_1\right) \otimes \left(\ket{\Delta x/2}_2+\ket{-\Delta x/2}_2\right)
\end{equation}
where the state labels indicate the relative distance from the original position of each particle.  Further, we take the standard static approximation (i.e. ignore wavefunction spread) and neglect the modelling of the opening and closing of the spatial superposition of the nanoparticles. For distinguishable superpositions --~i.e. $\Delta x$ much greater than the spread of the wavefunction in each branch -- the states $\ket{\pm \Delta x/2}_i$ are approximately eigenstates of their corresponding position operators $\hat{X}_i$, meaning that, effectively, each mass behaves as a qubit. This is the case in non-Gaussian matter-wave interferometry, where a qubit is used to drive, and measure, a large superposition~\cite{bose_spin_2017}.

In Appendix~\ref{app:dynamics}, we compute the time evolution of the center-of-mass degrees of freedom of the joint quantum system. Under the standard approximations we have outlined, the negativity is a necessary and sufficient condition for entanglement~\cite{horodecki_separability_1996,horodecki_quantum_2009}: $\lambda^{
\text{pt}}_-$ is the smallest, potentially negative eigenvalue of the partially transposed density matrix, which detects entanglement for $\lambda^{
\text{pt}}_-<0$. From its time evolution one can conclude that the state is never entangled, i.e. $\lambda^{\text{pt}}_- (t) \geq 0 \; \forall \; t $, if and only if 
\begin{equation}
\label{eq:nega_bound}
    \chi_{\textrm{GME}}=\frac{4\Gamma_{l}^2-\Gamma_{e}^2 }{\dot{\phi}_{e}^2 }\geq  1
\end{equation}
where
\begin{align}
\hspace{-0.3cm}\frac{\dot{\phi}_{e} [g] \kappa }{4M^2} &= -\int \frac{d^3k}{(2\pi)^3}\,\frac{|\rho_g(k)|^2}{k^2}\cos(k_z d)\sin^2\!\left(\frac{k_z\Delta x}{2}\right), \nonumber
   \\
 \frac{\Gamma_{l} [g, D]}{2 M^2}&=\int \frac{d^3k}{(2\pi)^3}\,
 |\rho_g(k)|^2 D(k) \sin^2 \left(\frac{k_z \Delta x}{2} \right),   \label{eq:rate_loc} \\
\frac{\Gamma_{e}  [g, D]}{ 4 M^2 }&=\int \frac{d^3k}{(2\pi)^3}\,
 |\rho_g(k)|^2 D(k)  \cos( k_z  d) \sin^2 \left(\frac{k_z \Delta x}{2} \right) . \nonumber
 \end{align}
These quantities describe the entangling phase ($\dot{\phi}_e$), the rates of local decoherence ($\Gamma_l$), and of the correlated decoherence  ($\Gamma_e$). The first depends functionally on the smearing function alone, while the latter two also depend on the noise kernel. This is the correlated-noise generalisation of the result obtained for local decoherence \cite{toros_relative_2021,schut_micronsize_2023,bose2026white}, which is recovered as a special case. Interestingly, the correlated decoherence assists the entanglement for fixed local rates. Recently, correlated decoherence effects, originating from vacuum fluctuations of the gravitational field~\cite{mitrakos_when_2026}, have been discussed in quantum gravity, as well as in a study that proposes noise measurements as an alternative to GME experiments to test Galilean-invariant theories of non-quantized gravity~\cite{minimal_fabiano_2026}. In particular, Ref.~\cite{minimal_fabiano_2026} derives an equivalent version of this no-entanglement condition -- although we retain the exact finite-separation expression for all the rates, and treat smearing explicitly. Their general framework covers several models of non-quantized gravity, while their explicit CQ analysis specialises to ultralocal decoherence and diffusion coefficients.

\subsubsection{Breaking up the entanglement threshold}

It is useful to break up the entanglement condition of Eq.~(\ref{eq:nega_bound}) into three parts encoding the noise, geometry, and smearing effects. We call $\{\Gamma_{l,*}, \Gamma_{e,*} , \dot{\phi}_{e,*}\}$ the set of coefficients computed for the DP kernel without smearing, namely, $\dot\phi_{e,*} \equiv \dot\phi_{e}[g =\delta]$  and $\Gamma_{i,*} \equiv \Gamma_{i}[g =\delta, D = D_*]$, with $i \in \{l,e\} $. Further, we define a dimensionless quantity encoding the amount of noise observed with respect to the minimal prediction without smearing
\begin{equation}
\label{eq:nu_exp}
    \nu_{\text{exp}}\equiv \frac{4 \Gamma_l^2 [g, D] - \Gamma_e^2  [g, D]}{4 \Gamma_{l,*}^2- \Gamma_{e,*}^2 } \ .
\end{equation}
Similarly,
\begin{equation}
    \mathcal{G}(d,\Delta x)\equiv \frac{C_*(0,\Delta x)}{|C_*(d,\Delta x)|} \ ,
    \label{eq:geo_fac}
\end{equation}
is a dimensionless number giving a geometric score to the experimental (bare) geometry, where
\begin{equation}
\begin{split}
 C_*(d,\Delta x)&\equiv\int \frac{d^3k}{(2\pi)^3}\,\frac{|\rho(k)|^2}{k^2}\cos(k_z d) \sin^2 \left(\frac{k_z \Delta x}{2} \right) \ ,
 \label{eq:C_d}
\end{split}
\end{equation}
In the small-displacement limit, the denominator of $\mathcal{G}$ is the geometric form factor used by Ref.~\cite{Tang:2024kdn} for two gravitationally coupled oscillators. Note that $\dot{\phi}_{e,*}=-4M^2 C_*(d,\Delta x)/\kappa$. Finally, we also introduce
\begin{equation}
    \epsilon_\phi\equiv \frac{\dot\phi_{e} [g]}{\dot\phi_{e,*} }-1 \ ,
\end{equation}
a factor encoding the deviation from the bare entanglement rate due to the smearing.

\begin{figure}[!t]
    \centering
    \includegraphics[width=0.95\linewidth]{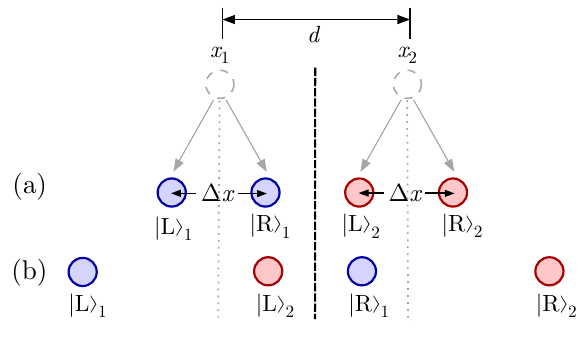}
    \caption{\label{fig:plane_sep} Collinear GME setup (a) respecting or (b)~violating the plane-separation condition.}
\end{figure}

These coefficients allow a particularly illuminating rewriting of the entanglement condition:
\begin{equation}
\label{eq:chi_GME}
    \chi_{\text{GME}}  = \nu_{\text{exp}} \frac{\mathcal{G}^2-1}{(\epsilon_{\phi}+ 1)^{2}}  \ ,
\end{equation}
suggesting that three effects -- deviations from the bare Newtonian potential, noise levels, and the geometry of the setup -- contribute to GME.

In fact, these quantities are not all independent, but they mainly capture the effects stated above. Being insensitive to the noise kernel and the UV regulator, the geometric factor $\mathcal{G}$ is only affected by the bare geometry of the setup. At fixed geometry, $\epsilon_\phi$ captures the effect of smearing, whilst $\nu_{\mathrm{exp}}$ captures the strength and correlations of the noise kernel -- which are mainly affected by changes in $D$, but are also sensitive to the smearing.

\subsubsection{Mutually exclusive conditions for entanglement}

Indeed, it is possible to identify three separate conditions, one for each contribution, that, when simultaneously true, rule out GME. This is the first technical result of our paper:
\begin{widetext}
\noindent\fbox{\begin{minipage}{0.965\textwidth}
\textbf{Theorem.}
Consider two identical compact matter distributions. Their centres of mass are placed at a distance $d$ and delocalised over $\Delta x$ in that same direction, along which their size is $R$. Let the evolution be described by Eq.~(\ref{eq:master}). Then, if the equal-superposition separable state $\ket{++}$ between the two masses evolves into an entangled state, these three conditions cannot be simultaneously true:
\begin{enumerate}
\item the gravitational field couples to the bare mass density: $g(x)=\delta(x)$
\item the decoherence-diffusion trade-off is satisfied: $D(k)\geq D_*(k)$ \;
\item the bare masses can be separated by a common plane perpendicular to their separation vector, in all branches of the superposition: $d-\Delta x \geq 2R$.
\end{enumerate}  
\end{minipage}}
\end{widetext}

We prove the theorem by contradiction. First, assume condition (1), such that $\epsilon_\phi = 0$. In the following argument, all quantities are therefore the bare, unsmeared ones. We further assume condition (2), the DD trade-off. Since the specific combination of the decoherence rates appearing in the entanglement threshold is functionally monotone under noise-kernel choices, in general 
\begin{equation}
    4\Gamma_l^2[D] - \Gamma_e^2[D] \geq 4\Gamma_l^2[D_*] - \Gamma_e^2[D_*] \ ,
\label{eq:nu_exp_bound}
\end{equation}
we have a lower bound for the LHS of Eq.~(\ref{eq:nega_bound}) (note that this is also in principle valid for arbitrary smearing). Hence, the DP kernel is the one that maximizes the generation of entanglement for any matter distribution:
\begin{align}
    4\Gamma_l^2-\Gamma_e^2 &\geq \left(\frac{4M^2}{\kappa}\right)^2 (C_*(0,\Delta x)^2-C_*(d,\Delta x)^2) \nonumber \\
    & \geq \dot{\phi}_e^2 \left[ \frac{C_*(0,\Delta x)^2}{C_*(d,\Delta x)^2} - 1 \right]  \ .
\end{align}
Then, no entanglement is generated if the following geometric condition is met
\begin{equation}
    \mathcal{G}(d,\Delta x)= \frac{C_*(0,\Delta x)}{|C_*(d,\Delta x)|} \geq\sqrt{2} \equiv \mathcal{G_*}\ ,
    \label{eq:geo_cond}
\end{equation}
essentially the no-entanglement condition of Eq.~(\ref{eq:nega_bound}) at the minimum noise kernel under the DD trade-off. We conclude the proof by exploiting an inequality derived in Appendix~\ref{app:common_plane}. If there exists a common plane that separates the two compact masses in each branch of the superposition, then the geometric factor is lower-bounded by $\mathcal{G}_\text{plane}(d,\Delta x) \geq 2 \geq \mathcal{G}_* $, proving that the two masses do not entangle.

\section{The price of entangling \\
when gravity is classical\label{sec:CQ}}

\begin{figure}[!t]
    \centering    \includegraphics[width=0.99\linewidth]{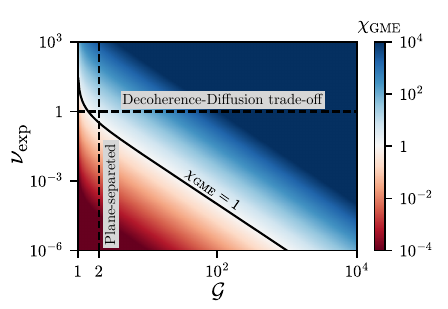}
    \caption{\label{fig:chi} Entanglement criterion $\chi_{\mathrm{GME}}$ for unsmeared masses ($\epsilon_\phi=0$), against the geometric factor ($\mathcal{G}$) and the experimental noise (relative to the minimal noise floor, $\nu_{\mathrm{exp}}$). The dark solid line is the entanglement threshold.}
\end{figure}

Consistent, linear CQ theories of gravity, in the Newtonian limit, satisfy the DD trade-off by construction -- a consequence of the dynamics being completely positive and trace-preserving. 
Moreover, restricting to the plane-separated geometry, depicted in Fig.~\ref{fig:plane_sep}, is not an \textit{ad hoc} condition. It is a common feature across many experimental proposals, due to the necessity of introducing a \emph{physical} plane to filter out all the non-gravitational interactions between the masses, and it is already used in current short-range gravity measurements~\cite{new_lee_2020,measurement_westphal_2021,van_de_kamp_quantum_2020,schut_relaxation_2023,schut_micronsize_2023}. Thus, the Theorem has straightforward, yet important consequence:
\begin{widetext}
\noindent\fbox{\begin{minipage}{0.965\textwidth}
\textbf{Corollary 1:}
For the class of Newtonian models described by Eq.~(\ref{eq:master}), classicality of the field is a sufficient condition for non-entangling dynamics between bare plane-separated masses, in the absence of smearing.
\end{minipage}}
\end{widetext}
In other words, GME detection would exclude any classical-quantum Newtonian theory without smearing, which cannot entangle notwithstanding of the non-locality of the noise kernel and the presence of the $1/\hat{r}$ effective Hamiltonian. One can see this explicitly in Fig.~\ref{fig:chi}, which, in the absence of smearing, visualizes interplay between noise and geometry for entanglement generation.

At first, this might seem to be in contradiction with explicit computations in the DP model~\cite{trillo_diosi_2025,entanglement_angeli_2025}. As anticipated, the solution to this tension relies on the smearing of the mass-density operator, first introduced in hybrid models to curb the problem of spontaneous heating (i.e. the injection of energy into the system at a large  rate~\cite{Ghirardi:1989bb,oppenheim2025diffusion}). As we shall see, the inclusion of this additional nonlocal scale allows the masses to entangle, but at a high price.

\begin{figure*}[t!]
    \centering
    \includegraphics[width=0.95\linewidth]{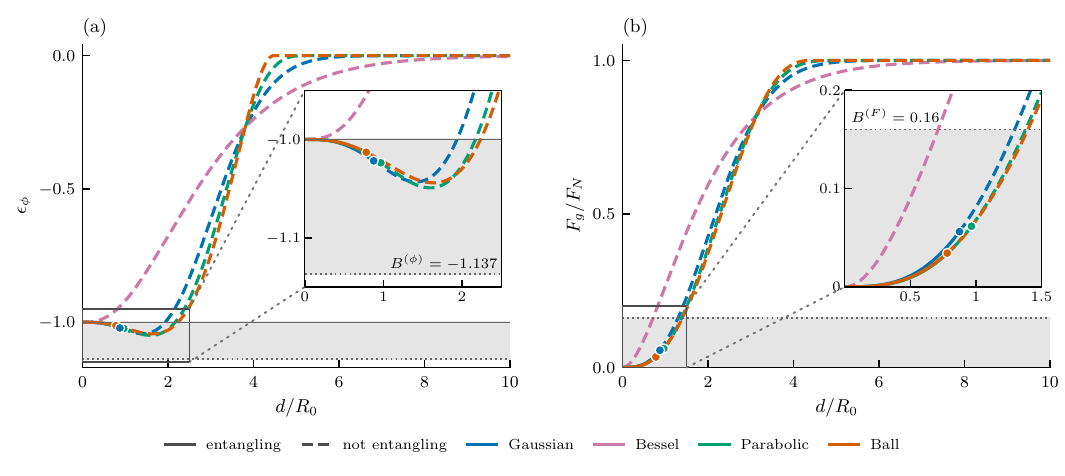}
    \caption{\label{fig:phase_force} (a) Phase-rate deviation $\epsilon_\phi$ (in the small-$\Delta x$ limit) and (b) smeared-to-Newtonian force ratio $F_g/F_N$ between two point masses against $d/R_0$, for the Gaussian, Bessel, parabolic and ball smearings. The solid lines indicate that entanglement under the DD trade-off is possible for some $\Delta x$ (Corollary~2), i.e.\ $d/R_0\leq\max_{\Delta x}d_\star/R_0$; the dahsed portion of the curves instead indicate the non-entangling regime. Circles mark the entanglement boundary, $d/R_0=\max_{\Delta x}d_\star/R_0=0.875$, $0.966$ and $0.782$ for the Gaussian, parabolic and ball smearings respectively; on the contrary, the Bessel smearing never entangles.}
\end{figure*}

Introducing smearing forces $\epsilon_\phi \neq 0$, opening the possibility for plane-separated masses to become entangled even with a classical mediator. However, this does not mean that GME detection would be inconclusive. In fact, this phenomenon comes with a significant change to the quantity that GME experiments aim to detect: the measured entangling phase differs from the expected one given by the unsmeared Newtonian kernel, i.e. $\dot\phi_e[g]\neq\dot\phi_{e,*}$ since, as discussed, smearing the mass density has an equivalent interpretation in terms of a UV regularisation of the Newtonian kernel. The deviation in the Newtonian kernel is not at all small -- we prove in Appendix~\ref{app:phase_rate_bound} that deviation of the phase rate $\epsilon_{\phi}$ is bounded, under the DD trade-off and for bare spherically-symmetric plane-separated geometries
\begin{equation}
    -1.137\approx B^{(\phi)}<\epsilon_{\phi}<-1 \ ,
\end{equation}
irrespective of the specific spherically-symmetric smearing function in the model. The exact integral expression for the phase bound $B^{(\phi)}$ can be found in Appendix~\ref{app:phase_rate_bound}. This indicates that not only is the phase rate suppressed in magnitude  by at least $86\%$ with respect to the Newtonian prediction, but its sign is also reversed.

Any witness that certifies entanglement is experimentally constructed via sets of correlated measurements -- e.g. heterodyne measurements of the nanoparticles' relative quadratures~\cite{serafini_quantum_2017,krisnanda2020observable,weiss_large_2021,fujita_inverted_2023,pedernales_enhancing_2022,poddubny_nonequilibrium_2025,toros_relativistic_2025} or spin correlations in hybrid qubit--nanoparticle systems~\cite{margalit_realization_2018,bose_qubit_2006,scala_matter_2013,romero_isart_quantum_2012}. These measurements would be able to distinguish the standard Newtonian prediction from the classical gravity mechanism -- especially through the sign reversal. Note, however, that this is a \emph{necessary} condition for GME with fundamentally classical gravity, not a sufficient one. Nonlocal modifications to quantum gravity can produce a similar suppression~\cite{biswas2012towards,Frolov:2015bia,sundrum2004fat}.

 There is a further, crucial consequence of this analysis, which affects observables far easier to measure than~$\dot{\phi}_e$. Take the simple example of Gaussian smearing. In that case the gravitational force between the two particles interpolates between the standard inverse-square law at large distances ($d\gg R_0 \implies F_{R_0}=G M^2 /d^2 $) and a linear behaviour at small distances ($d\ll R_0 \implies F_{R_0}=G M^2 \frac{d}{6\sqrt{\pi}R_0^3}  $). The boundary between these two asymptotic regimes is $d\sim R_0$, suggesting that GME is possible (although not guaranteed) only if there is a complementary modification to the law of gravitational attraction at the scales being probed. Indeed, in Appendix~\ref{app:force_bound} we explicitly derive the largest distance to observe entanglement, in the specific case of Gaussian smearing, to be $d_*\approx 0.83 R_0$ in the small $\Delta x$ limit -- in agreement with the computations of Refs.~\cite{trillo_diosi_2025,angeli_positivity_2025}. At that threshold, the ratio between the Newtonian prediction $F_N$ and the Gaussian-smeared force is
$    F_{R_0}(d_*)/F_N(d_*) \sim 0.05 $,
meaning that there must be approximately a 95\% suppression of the gravitational force between the masses, in addition to a functionally different dependence on $d$, and damping in time of the coherence.

Similarly to what happens to the phase rate, the strong suppression of gravitational attraction is not specific to Gaussian smearing. In Appendix~\ref{app:force_bound}, we extend this analysis to any positive, normalized, spherically symmetric smearing kernel. Thus, the main Theorem has the following crucial corollary:
\begin{widetext}
\noindent\fbox{\begin{minipage}{0.965\textwidth}
\textbf{Corollary 2:} 
No classical–quantum Newtonian theory described by Eq.~\ref{eq:master} can entangle masses whose bare configurations are plane-separated at distance $d$, unless this is accompanied by a violation of Newton's force law
\begin{equation}
    \frac{F [g] (d)}{F_N(d)}\leq B^{(F)} \approx 0.16 \ ,
\end{equation}
between spherical masses for any positive, normalized, spherically symmetric smearing $g$, and any $\Delta x$.
\end{minipage}}
\end{widetext}

The integral expressions for the bound $B^{(F)}$ can be found in Appendix~\ref{app:force_bound}. Note that the bound is independent of $\Delta x$ -- meaning that a classical precision-force experiment would measure a reduction of at least $84\%$ of the gravitational force compared to the Newtonian prediction at the scale where entanglement could be generated, for \emph{some} $\Delta x$. We stress that both the phase and force bound are not shown to be optimal -- and in fact we illustrate in Figure~\ref{fig:phase_force} that they are generous, at least for the set of smearing functions we consider.

This result provides model-independent force criteria to exclude the only mechanism by which these theories can entangle, under the standard plane-separated condition. In Section~\ref{sec:experiments} we discuss how existing and prospective short-range gravity measurements can probe a possible force suppression and consequently rule out such a mechanism. Note that, of course, quantum theories of gravity with UV regularisation of the Newtonian potential do not need to obey the force-suppression bound to entangle, since they do constrained by the DD trade-off.

Our discussion also admits a geometric interpretation. To understand the origin of entanglement purely geometrically, by defining the geometric factor for the \emph{smeared} masses
\begin{equation}
   \mathcal{G} [g] \equiv \frac{C[g](0,\Delta x)}{|C[g](d,\Delta x)|}
\end{equation}
where $C_{g}$ is defined as in Eq.~(\ref{eq:C_d}) with the smeared density distribution $\rho_g(k)=\rho(k)g(k)$, and
\begin{equation}
    \nu_{\mathrm{exp}}[g] \equiv \frac{4 \Gamma_l^2 [g, D] - \Gamma_e^2  [g, D]}{4 \Gamma_{l}^2[g, D_*]- \Gamma_{e}^2[g,D_*] }  \ ,
\end{equation}
the measured noise normalised with respect to the smeared noise floor. Although the new quantities cannot be computed from experimental observations alone, as they require knowledge of the fundamental smearing function itself (if it exists), they are easier to work with analytically. With these quantities, one can essentially reabsorb $\epsilon_\phi$, such that  
\begin{equation}
    \chi_{\mathrm{GME}} =\nu
_{\mathrm{exp}}[g] (\mathcal{G}^2[g]-1) \ .
\end{equation}

\begin{figure}[t!]
    \centering
    \includegraphics[width=0.95\linewidth]{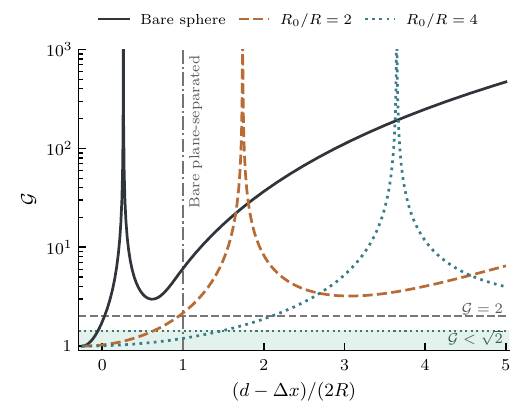}
    \caption{How the effective geometric factor $\mathcal{G}_g$ for compact spheres (solid line) is affected by Gaussian smearing (dashed lines). The green shaded region is the entangling region under the DD trade-off, which can be reached even for compact masses satisfying the bare plane-separation condition when smearing is introduced. The grey horizontal and vertical lines correspond to the lower bound of $\mathcal{G}$ on bare masses satisfying the plane-separation condition, and the plane-separation distance threshold, respectively. The “peaks” correspond to configurations in which the branches of the superposition accumulate relative phase at a negligible rate.}
    \label{fig:G_smeared}
\end{figure}

Under the DD trade-off, $\nu_{\mathrm{exp}}[g] \geq 1$ $\forall g(x)$, meaning that $\mathcal{G}[g]<\sqrt{2}$ is still required for entanglement. Note, however, that this time it is the \emph{smeared} geometric factor that needs to obey such a relation. The introduction of smearing evades our first theorem, i.e. lets bare plane-separated point masses entangle under the DD trade-off, precisely by allowing their smeared mass density to violate our key geometric bound, even when their bare geometric factor does not. Refs.~\cite{trillo_diosi_2025,entanglement_angeli_2025} work this out explicitly for the simplest, best-studied model: the DP equation, with Gaussian smearing. They point out that entanglement is observed as long as the two particles are separated by at most a fraction of the smearing scale $R_0$. In Appendix~\ref{app:force_bound}, we link precisely their entanglement threshold to the violation of the geometric bound through smearing. Crucially, it is not enough that the smeared masses overlap. This can be seen in action in Fig.~\ref{fig:G_smeared}. 

\section{Closing in on classical gravity:\\ short-range gravity measurements}
\label{sec:experiments}

The theorem we proved elucidates the important challenges for GME experiments. On the one hand, irrespective of its nature, the experiments require an extremely high level of isolation from noise -- in particular, one needs to violate the DD trade-off. On the other hand, the masses need to be sensitive enough to resolve the Newtonian interaction at short range, which can be achieved through a compact experimental geometry. With this in mind, we shall investigate what conclusions one can draw from existing experimental data on the parameter space of classical gravity. This analysis identifies future experiments  that represent meaningful milestones toward GME detection.

\subsection{Short-range gravity measurements\\ upper bound~$R_0$ \label{subsec:r0_exp}}

Considering the feeble nature of gravitational interactions, experiments performing gravity measurements at short range are among the most remarkable achievements of modern experimental science. No deviation from the inverse-square law has ever been observed down to $\sim 50 \mu$m in a fully classical experiment~\cite{new_lee_2020}, and the Newtonian interaction has been measured with sub-100-milligram sources of gravity~\cite{measurement_westphal_2021}. As previously discussed, this suggests an upper bound for the smearing parameter of gravity (as noted in Ref.~\cite{aspemeyer_quantum_2026} for Gaussian smearing). Generically, the smearing of the matter distribution changes the Newtonian potential of identical masses to 
\begin{equation}
    U [g] (r)=- G M^2 \int \text{d}^3 x  \text{d}^3 y  \frac{[\rho*g](x) [\rho*g](y)}{|x-y+r|}
\end{equation}
The smearing kernel is positive and normalised, meaning that it can only weaken the gravitational attraction. 

The shortest-range gravity measurement performed to date~\cite{new_lee_2020} involves two circular disk torsion pendula. The source mass is modulated by the rotation of one pendulum, which is engraved with an array of additional masses. A lattice of rotating planar objects lies well outside of the spherical-masses approximation that we have been discussing when deriving force deviations. It is conceivable that the effect of certain smearing choices might be softened when accounting for the response of such an experimental setup (specifically on its angular degrees of freedom), meaning that a universal bound on $R_0$ for arbitrary smearing functions cannot be easily established from their data. However, the explicit modification to the Newtonian potential studied in Ref.~\cite{new_lee_2020}, namely, the Yukawa correction
\begin{equation}
\label{eq:yukawa_pot}
    U_y(r)= -\frac{G M_1 M_2}{r}\left(1  + \alpha e^{-r/\lambda} \right) \ ,
\end{equation} 
is exactly reproduced by the smearing choice
\begin{equation}
\label{eq:yukawa_smearing}
    g(r)= \frac{K_1(r/R_0)}{2\pi^2R_0^2 r} \ ,
\end{equation}
where $K_1$ is the modified Bessel function of the second kind of order one and $\lambda = R_0$. This specifically recovers the modified potential for $\alpha = - 1$, a consequence of the fact that smearing of the masses can only suppress the attraction. Thus, the analysis of Ref.~\cite{new_lee_2020}, which integrates the Yukawa pair potential over the full experimental geometry, applies to this smearing without further modelling. Their 95\% confidence limit on the $\alpha=-1$ modification then gives $R_0 \lesssim 27 \,\mu$m~\cite{new_lee_2020}. Yet, it should be noted that such a smearing kernel does not entangle (see Figure~\ref{fig:phase_force}) and behaves quite differently compared to the rest of the possible choices considered in this work. 

It is possible to provide immediate upper bounds on the length scale~$R_0$ for a large class of spherically-symmetric smearing functions, if one restricts the analysis to simpler experimental setups. These are the classical experiments measuring the gravitational attraction between spherical masses, of which the one providing the strongest bound is that of Ref.~\cite{measurement_westphal_2021}, where the Newtonian inverse-square law between two approximately $90$ mg gold spheres was tested at an average distance of $4.15$mm with a torsion pendulum. The only caveat to provide such a bound is that one needs to have control on the leakage of the smearing at infinity -- otherwise one can engineer a kernel with most of the mass concentrated in the centre, and a slowly decaying tail that realises the desired second moment, without significantly changing the Newtonian interaction. Precisely for this reason, however, these fine-tuned kernels cannot entangle ($\epsilon_\phi \approx 0$) and will also have stringent lower bounds from heating experiments (the effective nonlocality scale will be much lower than $R_0$ itself).  

\begin{figure}
\centering\includegraphics[width=0.99\linewidth]{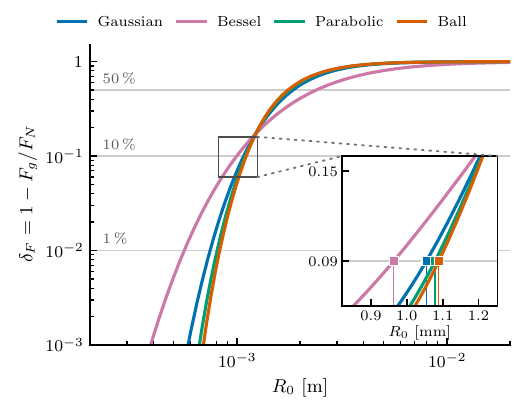}
    \caption{\label{fig:westpal} Fractional force deficit $\delta_F=1-F_g/F_N$ between two equal homogeneous spheres of radius $R=1.04$~mm at the average separation $d=4.15$~mm (Westphal set-up), against the smearing length $R_0$. Inset:  a force test with precision $\delta_F=9\%$ bounds $R_0\leq R_0^{\max}$ (filled squares) approximately given by: $1.05$, $0.96$, $1.08$ and $1.09$~mm for the Gaussian, Bessel, parabolic and ball smearings respectively.}
\end{figure}

An easy fix for this is to organise the smearing functions by their kurtosis, i.e. their fourth moment. In particular, the set of smearing functions whose fourth moment is constrained by
\begin{equation}
    \int d^3 x g(x) x^4 \leq c^4 R_0^4 \ ,
\end{equation}
is identified by the value $c$. Popular smearing choices (for example, all the ones in Figure~\ref{fig:phase_force}) satisfy this requirement for finite $c$ -- Gaussian smearing corresponds to $c= (15)^{1/4}\approx 2$. \footnote{Generically, any radial smearing with $g(r) \sim r^{-7-\epsilon}$ for large $r$ has finite kurtosis in three spatial dimensions if $\epsilon>0$.}  In Appendix~\ref{app:R0_bound}, we prove that, given an experiment that agrees with the Newtonian force prediction with sensitivity $\delta_{\textrm{exp}}$, one can bound 
\begin{equation}
    R_0 \leq \frac{d}{\sqrt{6-\sqrt{(2c^4+30)\delta_{\mathrm{exp}}}}}  \ .
\end{equation}
The functional form of the bound realises precisely its dependence on the fourth moment of the smearing function. In particular, whenever $c^4\geq 18/\delta_F-15$, no such bound even exists. Interestingly, however, for high sensitivity $(\delta_{\mathrm{exp}} \ll 1)$ the value of $c$ only enters at second order:
\begin{equation}
\frac{R_0^{\mathrm{max}}}{d} \approx \frac{1}{\sqrt{6}} + \frac{\sqrt{c^4+15}}{12 \sqrt{3}} \sqrt{\delta_{\mathrm{exp}}}  \ .
\label{eq:kurt_exp}
\end{equation}
Westphal et al.~\cite{measurement_westphal_2021} report a fit for Newton's constant $G$ approximately $9\%$ below the reference value, an offset consistent with their reported systematic uncertainties. We therefore adopt $\delta_{\mathrm{exp}}=0.09$ as a representative sensitivity and determine the smearing scale at which the predicted force deficit $\delta_F$ reaches this value, for smearing functions with controlled fourth moment. As numerically evaluated, the kurtosis dependence stays within $\pm 0.1$mm for the set of smearing functions considered, identifying $R_0^{\text{max}} \sim 1.7$mm, the leading order term in Equation~\ref{eq:kurt_exp}, as a conservative upper bound (see Fig.~\ref{fig:westpal}).

 \begin{figure}
\centering\includegraphics[width=0.99\linewidth]{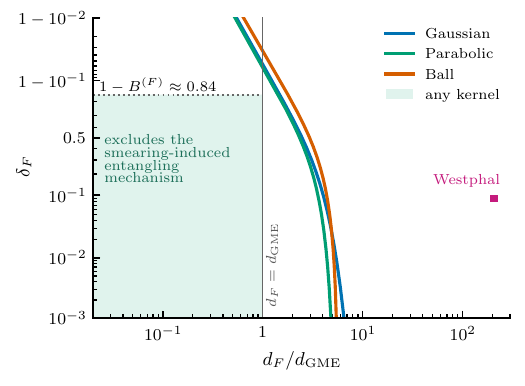}
    \caption{\label{fig:loophole}
    Required force precision $\delta_F$ of a force test that excludes any smearing-induced entangling mechanism of a GME experiment at separation $d_{\mathrm{GME}}$, against the experiment separation $d_F$ in units of $d_{\mathrm{GME}}$ (point masses). A test at separation $d_F$ with precision $\delta_F$ that lies below a smearing's curve excludes that smearing over this whole range; the solid curves plot $1-F_g/F_N$ at $d/R_0=(\max_{\Delta x}d_\star/R_0)\,d_F/d_{\mathrm{GME}}$ (for the illustrative Westphal marker, we used $d_{\mathrm{GME}} = 20 \mu$m as a proxy).}
\end{figure}

The analysis presented here has large consequences in the context of CQ theories of gravity. Spontaneous emission experiments~\cite{XENON:2025iov} lower-bound the Gaussian smearing scale to a few \r{A}ngstr\"om, beating the best lower bound attainable from unidentified acceleration noise in the LISA Pathfinder experiment~\cite{Helou:2016rii}. The inclusion of an upper bound from short-range gravity measurements allows the squeezing of the parameter space for the non-locality scale in classical theories from both sides. In the conservative case presented here, for instance, for the DP (Gaussian) smearing function, this is $5 \text{\AA}\lessapprox R_0 \lessapprox 1.7 \textrm{mm}$. It is reasonable to expect that, with an in-depth analysis of the results of~\cite{new_lee_2020}, the upper bound can be lowered significantly. 

\subsubsection{Excluding the smearing-induced entangling mechanism}

The distance $d$ in the proposed GME experiments is realistically limited to $d_{\mathrm{GME}} \gtrapprox 20\,\mu\text{m}$ by the required screening from unwanted Casimir forces~\cite{van_de_kamp_quantum_2020,schut_relaxation_2023,schut_micronsize_2023,bulling_stability_2026}. Can a future detection of entanglement at $d_{\mathrm{GME}}$ then be attributed unambiguously to non-classical gravity?

For the smearing-induced entangling mechanism to be a possible explanation of the phenomenon, the experiment would necessarily operate within the entangling range, $d_{\text{GME}} \leq d_*(R_0,\Delta x)$, and, by Corollary~2, the static attraction between the same masses at that separation would be strongly suppressed,
\begin{equation}
    \delta_F \equiv 1 - \frac{F[g](d_{\text{GME}})}{F_N(d_{\text{GME}})} \geq 1 - B^{(F)} \approx 0.84 \ .
    \label{eq:dF_GME}
\end{equation}
Since $\delta_F(d)$ can only grow at shorter separations, a short-range force measurement at any $d_F \leq d_\text{GME}$ that agrees with the Newtonian prediction within $\delta_\text{exp} < 1 - B^{(F)} \approx 0.84$ (i.e. less than 1$\sigma$ uncertainty) excludes a classical explanation of a potential GME observation. Remarkably, this requires only a force test of the inverse-square law, but at (or below) the very separation chosen for a GME experiment. Measurements at larger separations can exclude smearing-induced entangling mechanism only for specific choices of smearing function (see Fig.~\ref{fig:loophole}).

\subsection{Noise measurements \\
\& violating the DD trade-off \label{subsec:tradeoff}}
\begin{figure}[!t]
    \centering
    \includegraphics[width=0.95\linewidth]{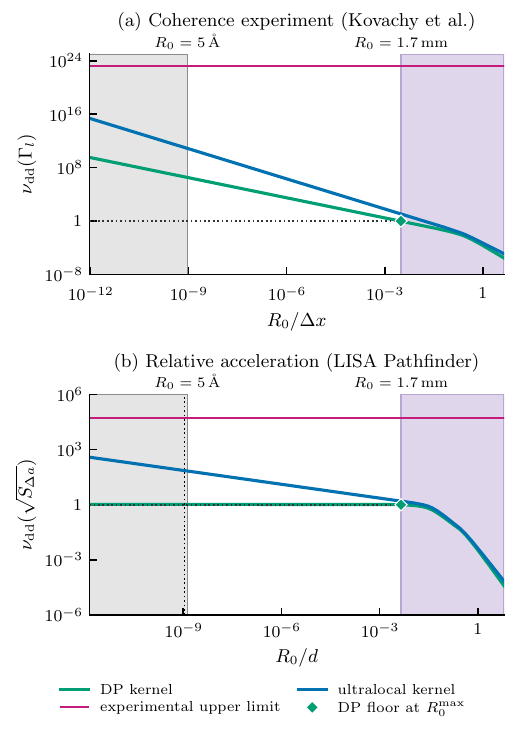}
    \caption{Distance from the violation to the DD trade-off floor, using the best bounds from a quantum and a classical experiment: (a) atomic-fountain experiments, bounding the local decoherence and (b) LISA Pathfinder, bounding relative acceleration noise. The blue line, for the ultralocal $\delta$-correlated kernel, represents the lower envelope allowed (i.e. minimising the further free parameter, $D_l$, pointwise on the plot for every $R_0$),  whilst the green line is the DP minimal noise kernel. The minimal floor is computed with the optimal smearing functions -- ball for (a) and parabolic for (b) in the range of allowed $R_0$. The shaded areas are the portions of the parameter range ruled out by spontaneous emission experiments~\cite{XENON:2025iov} (grey), and precision gravitational measurements~\cite{measurement_westphal_2021} (purple). The LISA pathfinder curves are computed by treating the masses as homogeneous rigid bodies -- this is justified by the fact that the intermolecular distance $d_{\text{inter}}$ (black dotted line) is smaller than the lower $R_0$ bound from spontaneous emission experiments.}
    \label{fig:noise_floors}
\end{figure}

Control and isolation of both classical and quantum systems have made it possible to reduce noise effects in a variety of experiments. This could probe the fundamental noises induced by the classicality of the gravitational field~\cite{Oppenheim:2022xjr,oppenheim2025diffusion,Angeli:2025ojs,Oppenheim:2026agp}: violating the DD trade-off would exclude all CQ hybrid theories of gravity outright. As we shall clarify, however, such an effort \emph{requires} an upper bound on $R_0$. The one derived in the previous section therefore provides a formidable ally in the quest to test the DD trade-off.

Consider a generic positive observable $O$, measured in the lab, that can be expressed as
\begin{equation}
    O[g, D] =\int d^3 k W_O(k) D(k) |\rho_g(k)|^2 \ ,
\end{equation}
with $W_O(k)\geq 0$. Important examples are
\begin{equation}
    W_{\Gamma_l}=M^2\left(1-\cos( k_z \Delta x)\right)
\end{equation}
for the local decoherence rates $\Gamma_l$, or
\begin{equation}
    W_{S_{\Delta a}}= 8k_z ^2
\sin^2\left(\frac{ k_z d}{2}\right),
\end{equation}
for the differential acceleration power spectrum ${S_{\Delta a}}$ (see below). 

Although the DD trade-off induces a lower bound on these observables in a fundamental theory of classical gravity through the DP kernel
\begin{equation}
     O[g, D]\geq O_{*}[g] =\int d^3 k W_O(k) D_*(k) |\rho_g(k)|^2 \ ,
\end{equation}
such a threshold is explicitly dependent on the smearing function $g$ (and hence on its scale $R_0$). Therefore, it can in principle be made arbitrarily low by increasing the nonlocality scale. At first, this seems a problem for the experimental accessibility of the DD trade-off violation (see Fig.~\ref{fig:noise_floors}, for how the DD floor changes as a function of $R_0$). However, as discussed, smearing always comes at the price of a modification of the gravitational force at short range, which can be used to lower-bound the minimal noise that this smearing procedure can achieve. 

Let $G_{\mathrm{exp}}$ be the set of smearing kernels in agreement with the force experiments. We define
$O_{\mathrm{min}}\equiv \inf_{\mathrm{g\in G_{\mathrm{exp}}}} (O_* [g])$, which becomes the relevant noise floor to beat in order to rule out the DD trade-off. In particular, given $R_0^{\textrm{max}}$, bounded from short-range gravitational experiments, $O_{\mathrm{min}}$ is the noise floor obtained with the choice of function $g = g^{\mathrm{opt}}$ which minimises the noise observable $O$, with $R_0 = R_0^{\mathrm{max}}$. If one is interested in the minimal observable for a specific decoherence kernel $D$, such as the ultra-local $D\propto \delta$ in Equation~\ref{eq:ultra_local_kerenl}, it suffices to perform the same minimisation replacing the DP kernel with the desired one.

Such a minimisation needs to be performed for each observable separately -- and curiously never seems to coincide with the Gaussian smearing kernel. For example, the optimal function to curb energy production has been explicitly shown in recent work~\cite{Piccione:2025zfb} to be a parabolic smearing. We have an in-depth discussion of optimal smearings for the observables of interest in Appendix~\ref{app:experimental}. In particular, we show that although never optimal, the analytically tractable Gaussian smearing is adequate as long as one is interested in order-of-magnitude estimates. 

In practice, there are two equivalent sets of experiments, one classical and one quantum in nature, that can help cross the DD trade-off. The original proposal to experimentally test the DD trade-off broke up the task into two: a classical experiment to bound the noise floor in the gravitational field directly, and a quantum one bounding the primary decoherence rate instead~\cite{Oppenheim:2022xjr}. Here, we focus on using the outcome of a \emph{single} experiment to reach the same result. This is a key improvement, since it removes the need to compare noise measurements from widely different setups. Moreover, that proposal did not characterise the key effects of changing the smearing length, which we now do in full detail.

An indirect way to measure the noise floor, and consequently probe the DD trade-off, is to measure the decoherence rate ($\Gamma_l$) of single masses in quantum superposition. For illustrative purposes, consider a Gaussian-smeared point particle (a good approximation in these types of experiments, that involve very small objects). The minimal decoherence rate interpolates between (reintroducing factors of $\hbar$)
\begin{equation}
\Gamma_{l,*} =
    \begin{cases}
        G M^2/(\sqrt{\pi} \hbar R_0)\ , \qquad &\Delta x \gg R_0 \\
GM^2\Delta x^2/(12\sqrt\pi\,\hbar R_0^3)
\ , \qquad &\Delta x \ll R_0
    \end{cases}
\end{equation}
showing a significant dependence on $R_0$. The comparison between the measured decoherence and the minimal noise gives the distance from the DD trade-off as 
\begin{equation}
    \nu_{\text{dd}} (\Gamma_l) = \frac{\Gamma_l}{\Gamma_{l,\mathrm{min}}} \;.
\end{equation}
Although the effect of the Gaussian smearing gives closed-form expressions that are easy to interpret and manipulate, in principle the optimal smearing function for the local decoherence is a uniform ball, as derived in Appendix~\ref{app:experimental}. Using the upper bound on $R_0$, one can then numerically compute the actual $\Gamma_{l,\mathrm{min}}$, which we report in Fig.~\ref{fig:noise_floors}.

For instance, the experimental parameters of the most coherent atomic-fountain experiment~\cite{quantum_kovachy_2015} ($\Delta x \approx 54$cm, $M=87$a.m.u.), which manage to keep coherent superpositions of $^{87}$Rb atoms up to $t\approx2.04$s, yield the illustrative $\nu_{\text{dd}} \approx 1.99 \times 10^{23}$ for Gaussian smearing -- see Figure~\ref{fig:noise_floors} for the value computed with the optimal ball kernel. More massive objects in superposition are found to be less stringent probes of the DD trade-off, as they tend to remain coherent for significantly shorter periods.
This is several orders of magnitude shy of the objective. Environmental noise in delicate quantum experiments is still a large limiting factor, not only in the ambitious path towards GME, but also in the test of the trade-off.

The most promising avenue to test the DD trade-off is, in fact, a purely classical experiment. Particularly strong are the experiments that monitor the effective random forces on two masses, with zero mean and covariance
\begin{equation}
    \mathbf{E}[F_a^i(t) F_b^j(t')]= Q^{ij}_{ab} \delta(t-t') \ ,
\end{equation}
where $Q_{ab}$ are empirical correlations between the forces acting on mass $a$ and $b$, whilst the $\{i,j\}$ indices indicate spatial directions. We discuss in Appendix~\ref{app:experimental} how to explicitly connect the empirical correlation matrix $Q$ with the fundamental noise kernel. The DD trade-off implies a minimal noise floor both in the acceleration power spectrum of a single particle and in the relative acceleration between two masses. 

LISA Pathfinder, a space experiment involving two cubic masses of side $L\approx 46$mm and average centre-of-mass separation $d\approx 376$mm, measured precisely this differential acceleration~\cite{Armano:2018kix}, finding the signal bounded by a minimal acceleration noise with power spectral density $S_{\Delta a} = 2\left[Q_{11}+Q_{22}-2Q_{12}\right]/M^2 = ( 1.7 \text{fm s}^{-2}\,\text{Hz}^{-1/2})^2$. Treating the masses as spherical, with radius $R= (3/4 \pi)^{1/3} L \approx 28.5$mm to preserve the actual mass and density of the object, and expanding to leading order in $R_0/R$ (reintroducing factors of $\hbar$):
\begin{equation}
S_{\Delta a, min}\approx \frac{4\hbar G}{R^3} \left[1-\frac{3R_0}{\sqrt\pi R} +\frac{2R_0^3}{\sqrt\pi R^3}\right] +\frac{8\hbar G}{d^3} \ ,
\end{equation}
we obtain the illustrative
\begin{equation}
  \nu_{\text{dd}} (\sqrt{S_{\Delta a}}) =    \frac{\sqrt{S_{\Delta a}}}{\sqrt{S_{\Delta a,\mathrm{min}}}}\approx 5.38\times 10^4 \ ,
\end{equation}
adopting Gaussian smearing. In Appendix~\ref{app:experimental} we show that the optimal smearing function for this observable is parabolic, for which we present the numerically computed floor in Fig.~\ref{fig:noise_floors}.

This result is striking, and deserves a comment. We are approximately five orders of magnitude from testing the crucial decoherence-diffusion trade-off through acceleration noise experiments, irrespective of the precise regularisation of the mass-density operator (if such a smearing exists at all). Closing the gap is possible in the near future, but it will require exquisite experimental efforts, other than detailed theoretical modelling, in order to extract stringent bounds on $R_0$.

\subsection{The gap from the entangling regime \label{subsec:milestone}}
Although violating the DD trade-off is a necessary condition for entanglement, it is not sufficient. Indeed, the relevant noise threshold for entanglement, even in a regime where smearing is negligible, is extremely sensitive to the geometric setup and is given by
\begin{equation}
\label{eq:nu_c}
    \nu_{\text{exp}}^c = \frac{1}{\mathcal{G}^2-1}\approx \frac{1}{\mathcal{G}^2} \ ,
\end{equation}
for which $\chi_{\text{GME}} = 1$. Let us consider typical GME experimental parameters -- two spherical bodies with mass $m\sim 10^{-15}$kg (implying $R \approx 0.4\,\mu$m for diamond density), $\Delta x \sim 10 \mu$m and $d\sim 3\times 10 \mu$m. This geometry corresponds to $\mathcal{G} \approx 7 \times 10^{2}$, implying that the critical noise below which entanglement is achievable is $\nu_{\text{exp}}^c \approx 2 \times 10^{-6}$. The experiment, then has to show a noise reduction with respect to the bare DD trade-off of approximately three orders of magnitude, since ($\nu_{\text{exp}}^c \approx \nu_{dd}^2 $). Yet, it should be clarified that, for this set up, the  acceleration noise suffices for GME is $\sim \text{fm}/\text{s}^2/\sqrt{\text{Hz}}$ \cite{grossardt_2020_acceleration,toros_relative_2021,bose2026white,braccini2026massindependentgravitationallyinducedentanglement}, as nanoparticles are not subjected to the LISA noise floor, which has been determined with much larger masses and the specific of that set up.

\subsubsection{Milestone experiments towards GME}
In this article, we have characterised in full generality a mechanism for GME generation in generic classical-quantum models of gravity in the Newtonian limit, the conditions in which such an effect is allowed (regularisation of the mass-density operator at the GME scale), and a physical smoking gun it would necessarily leave (a significant deviation from the inverse-squared law of gravity). This suggests the following roadmap towards and informative GME experiment: 
\begin{itemize}
    \item[(1)] \textbf{Excluding the smearing-induced entangling mechanism:} Perform a short-range gravity measurement to exclude a suppression of the Newtonian force $\delta_F < 0.84$ at the separation of future GME experiments ($d_{\text{GME}} \sim 20\mu$m).
    \item[(2)] \textbf{Exclusion of specific hybrid theories:} Lower the noise level observed in the experiment -- in either a single-body or composite system, quantum or classical -- below the floor provided by the chosen noise kernel. This can be done by beating the prediction with the optimal smearing function at the largest allowed nonlocal scale constrained by short-range gravity experiments ($R_0^{\text{max}}$).
    \item[(3)] \textbf{Exclusion of all continuous and Markovian classical theories:}  Lower the noise level of the experiment below the minimal DD trade-off floor, given by the optimal smearing choice with nonlocal scale $R_0^{\text{max}}$. 
    \item[(4)] \textbf{GME noise requirements:} If (1) and (3) are achieved,  lower the noise level in a two-body experiment further, below the critical threshold given by the geometry of the experimental setup: \mbox{$\nu_{\text{exp}}^c < 1/(\mathcal{G}^2 - 1)$}.
\end{itemize}

The first step (1) is required to exclude any deviation from Newton's inverse-square law at the probed scale, which could generate entanglement via the smearing mechanism. Steps (2)--(3) are necessary milestones that would progressively exclude hybrid theories of CQ gravity, up to the universal lower-bound on the noise floor. The final step encodes the noise isolation, given the geometry of the setup, that is necessary in order to generate entanglement: at that point, one can implement the entanglement witness for GME detection.

\section{Conclusions}
The presence of the $1/\hat{r}$-Newtonian operator interaction between two masses does not reveal the quantum or classical nature of the mediator. However, the presence of such a nonlocal operator alone does not imply that one would witness gravitationally-mediated entanglement (GME) also if gravity were fundamentally classical. Indeed, we prove that, in the plane-separated geometries relevant for the proposed experiments, continuous and Markovian classical-quantum (CQ) gravity in the Newtonian limit cannot generate entanglement, unless accompanied by a deviation ($\delta_F\gtrsim 0.84$) from the Newtonian inverse-squared law. Contextually, the smearing-induced mechanism for GME predicts also a significant suppression of the magnitude of the entanglement phase rate, and a sign flip with respect to the quantum gravity prediction, which would be manifest in appropriate witnesses of entanglement. Together, these observations restore the informative power of a potential witness of GME. The phase and force suppression bounds are model-independent within the class of dynamics we considered; that is, they hold for arbitrary noise kernels and smearing choices that respect Galilei symmetry. The key conceptual role of the plane-separation condition is to relax the necessity of assuming micro-causality to interpret the implications of an eventual entanglement detection~\cite{marshman_locality_2020}. Not only our geometric restriction of plane-separated masses has a clear interpretation of locality, but it also has experimental relevance, since physical Faraday shields are used in short-gravity measurements~\cite{new_lee_2020,measurement_westphal_2021}, and expected to feature in GME experiments.

Our results apply to the broad family of hybrid CQ theories that are Markovian and continuous. The no-entanglement theorem rests on a clear notion of locality and the requirement that Newton's law holds at the entangling distance. It is therefore natural to conjecture that these results will carry over to other theories of classical gravity in the enlarged Newtonian CQ class -- non-Markovian and jumping models. First, finite-sized collapse models reduce to the continuous class in the limit of small jumps~\cite{Pellegrini:2009mpr}. Similarly, a fundamental memory scale in a CQ model should be expected to be vanishingly small in the $c\to\infty$ limit, meaning that the Markovian regime should hold as an approximation universally. There are, of course, many exotic models in which gravity does not admit a quantum field description, but also has no classical interpretation (for example, the entangling Newtonian entropic gravity model presented in~\cite{Carney:2025wmz}). These fall outside the scope of our discussion. The Schr\"odinger-Newton model is not included in our analysis since it is nonlinear in the state, and it has already shown to be non-entangling~\cite{Struyve:2016obi,Fedida:2026xlp}

Inspired by the derived force-suppression theorem, we propose a method to upper-bound the smearing parameter of hybrid CQ theories of gravity for spherically symmetric smearing functions. Lower bounds have been extensively discussed (e.g.~\cite{XENON:2025iov,Helou:2016rii}), but a systematic discussion of upper bounds was lacking. Indeed, we illustrate how one can bound the nonlocality scale to $R_0^{\text{max}} \sim 1.7 $mm, by considering state-of-the-art short-range gravity experiment involving spherical masses. Our choice was conservative, but the constraint on $R_0$ can possibly be pushed further, in the spirit of the argument by Ref.~\cite{aspemeyer_quantum_2026}. Indeed, for a Bessel smearing function, the upper bound on the nonlocality scale can be placed already at the $\mu$m scale, by translating the bounds Ref.~\cite{new_lee_2020} gives on a specific modification to the Newtonian potential for non-spherical masses. Lifting this to a universal bound on the nonlocal parameter, however, requires careful modelling of the experimental setup used in such experiment -- it is conceivable that the effect of certain smearing choices is suppressed by the complicated geometric arrangement of masses used. With the explicit inclusion of a measurable upper bound, we highlight a new direction, on top of the minimal noise floor, that allows to squeeze CQ models from both sides -- and possibly rule them out altogether. As of today, there is no convincing evidence that establishes negligible smearing at the micrometre separations relevant to proposed GME experiments: this calls for further experimental investigations.

Leveraging the novel upper bound we derived, we show that there is convincing evidence that we are closing the gap towards an important milestone to reach GME -- testing the decoherence-diffusion trade-off. Indeed, we illustrate that the current noise sensitivity of LISA Pathfinder to differential acceleration noise is five orders of magnitude above the minimal noise floor of CQ models. This provides a reasonable target for the near future, with great discrimination significance -- violating the fundamental noise floor of CQ gravity would exclude \emph{all} the classical models of gravitation with continuous evolution. Reaching the required noise shielding to cover that ground is a formidable engineering task, but it is also, as we proved, a necessary step on the long path leading to a potential GME detection.

\section{Acknowledgments}
The authors would like to thank Daniel Carney, Giuseppe Fabiano, Flaminia Giacomini and Akira Matsumura for insightful discussions.
L.B. acknowledges support from the Engineering and Physical Sciences Research Council (EPSRC) through grants (EP/R513143/1 and EP/W524335/1) and from COST Action (CA23115). E.P. acknwoledges the support of the Italian Ministry of University and Research (MUR) under the Funding for Research Projects (FIS 2), pursuant to Ministerial Decree No. 23314 of 11/12/2024 (Project Q-GraSp, FIS-2023-02629). Further, EP acknowledges the support of the WOST, WithOut SpaceTime project (https://withoutspacetime.org), led by the Center for Spacetime and the Quantum (CSTQ), and supported by Grant ID\#~63683 from the John Templeton Foundation (JTF). The opinions expressed in this work are those of the author(s) and do not necessarily reflect the views of the John Templeton Foundation or those of CSTQ. A.S. acknowledges financial support from the Leverhulme Trust through Research Project Grant No. RPG-2024-287. SB would like to acknowledge EPSRC grant EP/X009467/1. This work was made possible through the support of the WOST, WithOut SpaceTime project (https://withoutspacetime.org), supported by Grant ID\#63683 from the John Templeton Foundation (JTF). S.B’s research is funded by the Gordon and Betty Moore Foundation through Grant GBMF12328, DOI 10.37807/GBMF12328, and the Alfred P. Sloan Foundation under Grant No. G-2023-21130. The ideas and proofs in this paper originated and were developed by the authors. Opus 5.5 and GPT 6 Astra were used in numerical evaluations, adversarial proof verification, some of the steps towards the force and phase bounds Eqs. 21 and 22, and editing of the draft. All content is the responsibility of the authors.

\appendix
\setcounter{secnumdepth}{1}
\onecolumngrid

\section{Gravitational Two-Body Effective Dynamics with Noise}
\label{app:dynamics}

We aim to describe the interaction between two masses in a quantum gravity theory, but with a noisy environment for both the Newtonian gravitational field and the matter distribution. This is performed by first integrating out the Newtonian field and then considering the effective action of the centre of mass (CoM) of the matter distribution. 

\subsection{Closed-time-path action and field elimination: Effective Action of the Matter Distributions}

We offer an intuitive derivation, in terms of open-system path integrals, of the dynamical map in Equation~\ref{eq:master} -- agnostic about the quantum or classical nature of the gravitational field. An illuminating derivation of the same equation in the language of measurement-and-feedback trajectories in classical-quantum models can be found in~\cite{Tilloy:2015zya}.

The gravitational action for a Newtonian potential $\Phi$ coupled to total matter density $m(x)= m_1(x)+m_2(x)$  given in Eq.~(\ref{eq:newton-action}), from which one can consider 
\begin{align}
    \rho(\Phi_f^+, \Phi_f^-, m^+_f, m^-_f, t_f) &= \mathcal{N} \int \mathcal{D} \Phi^+\mathcal{D} \Phi^-  \mathcal{D} m^+  \mathcal{D} m^- e^{I_{N} [\Phi^+, \Phi^-, m^+, m^-, t_i, t_f] } \rho(\Phi_i^+, \Phi_i^-,  m^+_i, m^-_i, t_i) \;, 
\end{align}
where, the closed time path the unitary action is
\begin{equation}
I_{N}^{U} = -i \left( S_N[\Phi_+, m^+] - S_N[\Phi_-, m^-]+ S_m[m^-]- S_m[m^+] \right)\,,
\end{equation}
where $+/-$ label the forward and backward time-evolution branches with addition of noise,  $S_m$ is the matter-only action containing, for example, the kinetic term of the matter field and any non-gravitational interaction and
\begin{equation}
 S_N[\Phi,m]=\int d t\,d^3 x\,
  \left[\frac{\kappa}{2}\left(\nabla\Phi(x)\right)^2+m(x)\Phi(x)\right] ,
  \label{eq:newton-action} 
\end{equation}

We now introduce the Keldysh rotation $\bar{\Phi} = (\Phi_+ + \Phi_-)/2$, $\Phi_\Delta = \Phi_+ - \Phi_-$, and similarly $\bar{m} = (m^+ + m^-)/2$, $m_\Delta = m^+ - m^-$. This basis for the path integral is often referred to as the “classical-quantum” rotation as well: the barred quantities encode the expectation values of the operators, and are hence sometimes called the classical variables; the differenced variables encode the quantum fluctuations instead. In this basis, the kinetic term reads
\begin{equation}
I_{N}^{U} \supset- \int d^4x \, \frac{i\kappa}{2}\left[(\nabla\Phi_+)^2 - (\nabla\Phi_-)^2\right] = -\int d^4x \,i\kappa\,\nabla\bar{\Phi}\cdot\nabla\Phi_\Delta = \int d^4x \,i\kappa\,\Phi_\Delta\nabla^2\bar{\Phi}
\end{equation}
after integration by parts. The matter-field coupling is
\begin{equation}
I_{N}^{U} \supset-i \int d^4x \, \left(m^+\Phi_+ - m^-\Phi_- \right)=-i\int d^4x \left(\, m_\Delta\,\bar{\Phi} + \bar{m}\,\Phi_\Delta\right)
\end{equation}
Thus, the action can be expressed as
\begin{align}
\label{eq:S_CTP}
I_{N}^U &=-i \int d^4x \, \left[\Phi_\Delta\! \left(-\kappa\nabla^2\bar{\Phi} + \bar{m}\right) + m_\Delta\,\bar{\Phi}\right]
\end{align}
In this form, it is trivial to include noise due to interaction with some environment (e.g. particles whizzing through the experimental apparatus, thermal noise). We include it both in the gravitational field and in the matter evolution, describing it through the noise spectra $W$ and $\lambda$, respectively. In short, we assume that the environments responsible for fluctuations, $\eta$ and $\zeta$, couple linearly to the system and have already been integrated out (and the Markovian limit has been taken). More precisely, we assume zero-mean Gaussian fluctuations in the Newtonian constraint and in the matter sector:
\begin{equation}
\langle \eta(x,t)\eta(x',t')\rangle = \frac{W(x,x')}{4}\delta(t-t'), \qquad \langle \zeta(x)\zeta(x')\rangle = \lambda(x,x')\delta(t-t'),
\end{equation}
with no cross-correlation between $\eta$ and $\zeta$. Then, the action for the open-system path integral is given by
\begin{align}
\label{eq:S_CTP}
I_{N} &= I_{N}^U - \frac{1}{2}\int dt\,  d^3x \, d^3x' \, \left( \Phi_\Delta(x) \frac{W(x,x')}{4}\Phi_\Delta(x') + m_\Delta(x) \lambda(x,x') m_\Delta(x') \right) \ .
\end{align}
By doing so, we have implicitly assumed that the state of the fluctuations around the Newtonian potential 
\begin{equation}
    \eta^\pm \equiv -\kappa \nabla^2\Phi^\pm + m^\pm \ .
\end{equation} 
factorises from the state of matter, and it is Gaussian, with zero mean and covariance function $W$.  At this point, it is worth pausing to highlight that the path integral so defined is precisely the path integral of a classical-quantum theory. By this, we mean that by reversing the Keldysh rotation in the matter sector, and performing the $\Phi_\Delta$ integration, one arrives precisely at the formulation of hybrid dynamics in the path-integral formalism ~\cite{oppenheim2023covariant}, a useful analogy at Gaussian order~\cite{Panella:2025sig}. We will rely on it to be formally agnostic about whether the noise in the gravitational field is due to the interaction with some environment or to fundamental irreversibility of the dynamics (necessary whenever some fundamentally classical degree of freedom backreacts onto a quantum system~\cite{oppenheim2022the-two-classes,galley_no-go_2022}).

In order to derive the effective action for the matter distribution, we can now integrate over both $\Phi_\Delta$ and $\bar{\Phi}$. 
\begin{align}
     \rho(m^+_f, m^-_f, t_f) &\propto \int d\bar\Phi_f\,\delta(\Phi_{\Delta f}) \int \mathcal{D} \bar{\Phi}\mathcal{D} \Phi_\Delta  \mathcal{D} \bar{m}  \mathcal{D} m_{\Delta} e^{I_{N}} \rho(\bar{m}_i, m^{\Delta}_i, t_i)  \rho_f(\nabla^2\bar{\Phi}_i-\bar{m}_i, \nabla^2\Phi_i^\Delta-m^{\Delta}_i, t_i) \;\\
     &\propto \int  \mathcal{D} \bar{m}  \mathcal{D} m_{\Delta} e^{I_{N}^{\textrm{eff}}} \rho(\bar{m}_i, m^{\Delta}_i, t_i) \ ,
\end{align}
with
\begin{equation}
    e^{I_{N}^{\textrm{eff}}} \equiv \mathcal{N'} \int d\bar\Phi_f\,\delta(\Phi_{\Delta f})\int \mathcal{D} \bar{\Phi}^+\mathcal{D} \Phi_\Delta \  e^{I_{N}} \rho_f(\nabla^2\bar{\Phi}_i-\bar{m}_i, \nabla^2\Phi_i^\Delta-m^{\Delta}_i, t_i) \ ,
\end{equation}
for the appropriate normalisation factor $\mathcal{N}'$, and $\rho_f$ the initial state of the fluctuations around the Poisson equation. The $\delta$-function factor comes from the trace condition on bra and ket, i.e. $\Phi_f^+ = \Phi_f^-$.

We do the integration in momentum space (such that $\nabla^2 \mapsto - k^2$). Integration over the average field $\bar{\Phi}$ results in the following constraint
\begin{equation}
    \delta(\Phi_\Delta +m_\Delta/\kappa k^2)
\end{equation}
which makes the integration over $\Phi_\Delta$ straightforward as well, leading to
\begin{equation}
\label{eq:S_CTP}
I_{N}^{\mathrm{eff}} 
= \int \frac{d^3k}{(2\pi)^3} \int dt \left[\frac{i}{2\kappa k^2} \left(\bar{m}(-k) m_\Delta(k) + m_\Delta(-k) \bar{m}(k) \right) - \frac{1}{2}m_\Delta(-k)\left(\frac{W(k)}{4\kappa^{2} k^4} +  \lambda (k) \right)  m_\Delta(k)  \right] \ .
\end{equation}
as claimed in the main text. 

The closed-time-path integral for matter described by the non-relativistic field $m(x)$ interacting linearly with the Newtonian potential $\Phi(x)$ is, in the average-difference basis $\bar f=(f^++f^-)/2$ and $f_\Delta=f^+-f^-$ (with $\pm$ indicating the bra and ket fields)
\begin{equation}
    \rho_t = \int \mathcal{D}\Phi \int \mathcal{D}m  \ e^{I_{QG}} \rho_0 \ .
\end{equation}
Here, $\rho_t$ is the state of the combined matter+gravity system at time $t$, and the functional integration is over both the average and difference fields, with appropriate boundary conditions. The action for the system is
\begin{align}
 I_{\rm QG}={}&-i\int d t\, d^3 k\,
 \left[\Phi_\Delta(- k)(\kappa k^2\bar\Phi( k)
 +\bar m( k))+m_\Delta(- k)\bar\Phi( k)\right]
 \nonumber\\
 &-\frac{1}{2}\int d t\, d^3 k\,
 m_\Delta(- k)\lambda(k)m_\Delta( k)
 \nonumber\\
 &-\frac{1}8\int d t\, d^3 k\,
 \Phi_\Delta(- k)W(k)\Phi_\Delta( k) \ ,
 \label{eq:CTP}
\end{align}
where the first line corresponds to the unitary gravitational action in the in-in formalism, whilst the second and third lines encode the covariance of the Gaussian noise in the matter and gravitational fields respectively.

The path integral being Gaussian in $\Phi$, the Newtonian field can be integrated out exactly, enforcing
$\Phi_\Delta=-m_\Delta/(\kappa k^2)$. The effective matter exponent is then
\begin{align}
 I_{\rm eff}=\int d t\, d^3 k\,
 \biggl[&\frac{i}{2\kappa k^2}
 \left(\bar m(- k)m_\Delta( k)
 +m_\Delta(- k)\bar m( k)\right)\nonumber\\
 &-\frac{1}{2}m_\Delta(- k) D(k)m_\Delta( k)\biggr] \ .
 \label{eq:S_eff_app}
\end{align}

This action has an equivalent description as a Lindbladian generator
\begin{equation}
    \partial_t\rho = -i[H_N,\rho]-\frac{1}{2}\int \frac{d^3k}{(2\pi)^3} D(k) \left[m(-k),[m(k),\rho]\right] \ ,
\end{equation}
\begin{equation}
    H_N = -\frac{1}{2\kappa}\int \frac{d^3k}{(2\pi)^3}\,
\frac{ m(- k) m( k)}{k^2},
\qquad \kappa=\frac{1}{4\pi G} 
\end{equation}
is the $1/\hat{r}$ Newtonian interaction (after dropping the irrelevant self-interaction terms), while the last term is the decoherence of the matter degrees of freedom. This is precisely the Lindbladian in the main text -- upon smearing of the mass-density operator.

\subsection{Effective Dynamics of the Centre of Mass}

We now review the standard qubitisation approximation for GME experiment~\cite{bose_spin_2017}, from the path-integral perspective. In order to describe the interaction between two masses, let us consider two general matter distributions $m^{\pm} (x) = m_1^{\pm}(x) + m_2^{\pm} (x) $. Each of the non-relativistic mass is superposed in with respect of its centre of mass $X_{A}^{\pm}$ and $X_{B}^{\pm}$, while we assume that their distributions are fixed and identical, i.e. $m^{\pm}_{i} (x) = M_i \rho(x - X_i^{\pm}) $, where $\rho$ is normalised, $\int \text{d}^3 x \rho (x) = 1$ and $M_i$ is the total mass of each particle. At the level of the path integral and of the measures $ \mathcal{D} m^\pm$, the assumption can be included as a constraint via 
\begin{equation}
    1 = \int \mathcal{D} X^{\pm}_1 \delta \left( m^{\pm}_1 (x) -  M_1  \rho(x - x^{\pm}_1) \right)  \int \mathcal{D} X^{\pm}_2 \delta \left( m^{\pm}_2 (x) -  M_2  \rho(x - x^{\pm}_2) \right)  
\end{equation}
such that one can trivially compute the matter-distribution integral and arrive at the effective action of the centre-of-mass motion
\begin{align}
    \rho\left (X_{1,f}^\pm, X_{2,f}^\pm, t_f \right) &= \mathcal{N} \int \mathcal{D} X_1^+\mathcal{D} X_1^-   \mathcal{D} X_2^+\mathcal{D} X_2^- e^{I_{N}^{\mathrm{eff}} [X_1^+, X_1^-, X_2^+, X_2^-,  t_i, t_f] } \rho\left (X_{1,i}^\pm, X_{2,i}^\pm, t_i \right) \;, 
\end{align}
where $f$ and $i$ label the final and initial boundary conditions, respectively, and integrating over $m_{i}^\pm (x)$ implies, via the delta function, $m^{\pm}_{i} (x) \to M_i \rho(x - X_i^{\pm})$. In momentum space, considering only the centre-of-mass degrees of freedom in the $z$-direction, the mass distributions take the general form $m^{\pm}_{i} (k) = M e^{-ik x^{\pm}_{i}} \rho(k) $. Thus, the action $ I_{N}^{\mathrm{eff}} [X_1^+, X_1^-, X_2^+, X_2^-,  t_i, t_f] $ can be computed by \textcolor{blue}{Eq.~(\ref{eq:S_eff_app})} via
\begin{align}
\bar{m} (k) &= \frac{M \rho(k)}{2} \left(   e^{-i k_z X^+_1} +  e^{-i k_z X^+_2}  +   e^{-i k_z X^-_1}  + e^{-i k_z  X^-_2 }  \right) \\
    m_\Delta (k) &= M \rho(k) \left(  e^{-i k_z X^+_1} +  e^{-i k_z X^+_2}  -  e^{-i k_z X^-_1}  - e^{-i k_z  X^-_2 }  \right)
\end{align}
such that 
\begin{equation}
\bar{m}^* m_\Delta + m_\Delta \bar{m}  = 2 M^2 |\rho(k) |^2  \left[ \cos (k_z (X_1^+ - X_2^+))-\cos (k_z (X_1^- - X_2^-) \right]
\end{equation}
\begin{equation}
    m_\Delta (k) m^*_\Delta (k) = M^2 |\rho(k) |^2  \left(  e^{-i k_z X^+_1} +  e^{-i k_z X^+_2}  -  e^{-i k_z X^-_1}  - e^{-i k_z  X^-_2 }  \right) \left(  e^{i k_z X^+_1} +  e^{i k_z X^+_2}  -  e^{i k_z X^-_1}  - e^{i k_z  X^-_2 }  \right) \nonumber
\end{equation}

Under the assumptions of static non-overlapping states, the states $\ket{\pm\Delta x/2}_i$ are eigenstates of $\hat{X}_i$. Given that no transition is allowed between these states during the dynamics, one can compute the time evolution of the initial state $\ket{\psi} = \left(\ket{\Delta x/2}_1+\ket{-\Delta x/2}_1\right) \otimes \left(\ket{\Delta x/2}_2+\ket{-\Delta x/2}_2\right)/2$ in the basis of eigenstates $\ket{j,k}$ with $j,k\in \{\pm \Delta x/2\}$, such that one finds the time evolution of the density matrix 
\begin{equation}
\label{eq:state_new}
\varrho_t=\frac{1}4
\begin{pmatrix}
 1 & e^{-\Gamma_l t + i(\dot{\phi}_l - \dot{\phi}_g^+)t} & e^{-\Gamma_l t + i(\dot{\phi}_l - \dot{\phi}_g^-)t} & e^{-(\Gamma_e + 2\Gamma_l)t} \\
 e^{-\Gamma_l t + i(\dot{\phi}_g^+ - \dot{\phi}_l)t} & 1 & e^{-(2\Gamma_l - \Gamma_e)t + i(\dot{\phi}_g^+ - \dot{\phi}_g^-)t} & e^{-\Gamma_l t + i(\dot{\phi}_g^+ - \dot{\phi}_l)t} \\
 e^{-\Gamma_l t + i(\dot{\phi}_g^- - \dot{\phi}_l)t} & e^{-(2\Gamma_l - \Gamma_e)t + i(\dot{\phi}_g^- - \dot{\phi}_g^+)t} & 1 & e^{-\Gamma_l t + i(\dot{\phi}_g^- - \dot{\phi}_l)t} \\
 e^{-(\Gamma_e + 2\Gamma_l)t} & e^{-\Gamma_l t + i(\dot{\phi}_l - \dot{\phi}_g^+)t} & e^{-\Gamma_l t + i(\dot{\phi}_l - \dot{\phi}_g^-)t} & 1
\end{pmatrix}
\end{equation}
where the rates $\Gamma_l$ and $\Gamma_e$ are given in Eq.~(\ref{eq:rate_loc}), while the rates of the phases are  
 \begin{align}
 \dot{\phi}_l&= \frac{M^2}{\kappa} \int \frac{d^3k}{(2\pi)^3}\,
 \frac{|\rho(k)|^2}{k^2}  \cos( k_z d )  \hspace{1cm}
 \dot{\phi}_g^{\pm} = \frac{M^2}{\kappa} \int \frac{d^3k}{(2\pi)^3}\,
 \frac{|\rho(k)|^2}{k^2}  \cos( k_z (d \pm \Delta x) ) 
\end{align}
and, integrating over the angular components, one finds the rates
\begin{align}
 \Gamma_l&=\frac{M^2}{2\pi^2} \int d k_r \,
 |\rho(k_r)|^2 D(k_r) k_r^2 \left( 1 - \frac{\sin( k_r \Delta x )}{k_r \Delta x} \right) ,\nonumber\\
 \Gamma_e
 &=\frac{M^2}{2\pi^2} \int d k_r \,
 |\rho(k_r)|^2  D(k_r) k_r^2
 \left[ \frac{2\sin(k_r d)}{k_r d}
  - \frac{1}{k_r}\left(\frac{\sin\!\big[k_r(d+\Delta x)\big]}{d+\Delta x} + \frac{\sin\!\big[k_r(d-\Delta x)\big]}{d-\Delta x}\right) \right]
\end{align}

\section{Negativity Bounds}
\label{eq:negativity}

The potentially negative eigenvalue of the partial transpose of Eq.~(\ref{eq:state_new}) is
\begin{equation}
\lambda^{\text{pt}}_- (t) = \frac{1}{4} \left[ 1 - e^{-2\Gamma_l t}\cosh(\Gamma_e t)
 - \sqrt{e^{-4\Gamma_l t}\sinh^2(\Gamma_e t)
 + 4e^{-2\Gamma_l t}\sin^2\!\left[ \dot{\phi}_{e} t / 2\right]} \right]
\end{equation}
where we defined the entangling phase rate
\begin{align}
    \dot{\phi}_{e} &= \dot{\phi}_g^+ + \dot{\phi}_g^- - 2\dot{\phi}_l
    = \frac{2M^2}{\kappa}\int \frac{d^3k}{(2\pi)^3}\,\frac{|\rho(k)|^2}{k^2}\cos(k_z d)\big[\cos(k_z\Delta x)-1\big] \\
    &= \frac{M^2}{2\pi^2\kappa}\int_0^\infty dk\,|\rho(k)|^2
    \left[\frac{\sin\!\big(k(d+\Delta x)\big)}{k(d+\Delta x)}+\frac{\sin\!\big(k(d-\Delta x)\big)}{k(d-\Delta x)}
    -\frac{2\sin(kd)}{kd}\right]
\end{align}

One can note that the PPT condition for separability $\lambda^{\text{pt}}_- (t)\geq0$ is equivalent to
\begin{equation}
 \cosh(2\Gamma_lt)-\cosh(\Gamma_et)
 \geq 2\sin^2(\dot{\phi}_et/2).
 \label{eq:ppt}
\end{equation}
For all $t\geq0$,
\begin{align}
    \cosh(2\Gamma_lt)-\cosh(\Gamma_et) \geq (4\Gamma_l^2-\Gamma_e^2)\frac{t^2}{2} \geq \dot{\phi}_e^2\frac{t^2}{2}\geq 2\sin^2(\dot{\phi}_et/2)
\end{align}
where the first inequality relies on $2|\Gamma_l|\geq \Gamma_e$, whilst the penultimate follows from
$4\Gamma_l^2-\Gamma_e^2\geq \dot{\phi}_e^2$. This also follows from expanding at linear order in time: the entanglement rate is 
\begin{equation}
    \left.\frac{\mathrm{d} \lambda^{PT}_-}{\mathrm{d}t}\right|_0 =\frac{1}{4} \left(2\Gamma_l - \sqrt{\Gamma_e^2 +  \phi_g^2} \right)
\end{equation}
implying that for $2\Gamma_l <  \sqrt{\Gamma_e^2  +  \phi_g^2} $ entanglement is generated. Given that for two qubits the negativity is a necessary and sufficient condition for entanglement ($R_e < 0$), we shall focus on the converse, i.e. entanglement is not generated for $2 \Gamma_l \geq  \sqrt{\Gamma_e^2  +  \phi_g^2} \implies 4\Gamma_l^2 \geq  \Gamma_e^2  +  \phi_g^2 \implies  4\Gamma_l^2  -  \Gamma_e^2  \geq   \phi_g^2 $ as, for $W(k)>0 $ and $\lambda(k)>0$, $\Gamma_l > 0$. 

\section{Common-plane separation implies the geometric bound}
\label{app:common_plane}

We prove that the geometric condition required in the main text is automatically satisfied whenever all branches of the two bodies can be separated by a common plane. We consider identical rigid bodies and take the direction normal to the separating plane to be the $z$ direction, which is also the direction of the spatial superpositions.

Let $\rho(x)$ be the normalized mass profile of either body, and assume that its support is contained in a slab of width $L$ along $z$. The two branches of each body are separated by $\Delta x$, while the mean separation between the bodies is $d$. The condition
\begin{equation}
    d-\Delta x \geq L
    \label{eq:plane-separation}
\end{equation}
ensures that one plane separates both branches of the first body from both branches of the second body.

The key object we want to bound is the kernel-weighted material overlap
\begin{equation}
    C_*(d,\Delta x) =\int \frac{d^3k}{(2\pi)^3}\, \frac{|\rho(k)|^2}{k^2} \cos(k_zd) \sin^2\left(\frac{k_z\Delta x}{2}\right) \ ,
    \label{eq:Cstar-definition}
\end{equation}
which can be rewritten as
\begin{equation}
    C_*(d,\Delta x) = \frac{2J(d)-J(d+\Delta x)-J(d-\Delta x)}{4} \ ,
    \label{eq:Cstar-J}
\end{equation}
where we defined
\begin{equation}
    J(d) = \int \frac{d^3k}{(2\pi)^3}\,\frac{|\rho(k)|^2}{k^2} \cos(k_zd) \ .
    \label{eq:J-definition}
\end{equation}
In particular, we will prove that, if there exists one plane that separates both branches of the two bodies, then the stronger bound
\begin{equation}
    \left|C_*(d,\Delta x)\right| \leq \frac{1}{2}C_*(0,\Delta x) 
    \label{eq:strong-geometric-bound}
\end{equation}
applies, so that the threshold in the main theorem is automatically satisfied.

To prove Eq.~\eqref{eq:strong-geometric-bound}, we Fourier-transform only in the directions $x_\parallel$ parallel to the separating plane:
\begin{equation}
    \rho_{q}(z) = \int d^2x_\parallel\, e^{-iq\cdot x_\parallel} \rho(x_\parallel,z) \ ,
    \label{eq:partial-Fourier}
\end{equation}
with the subscript indicating the magnitude of the wavevector. Further, it is useful to define the partial Fourier transform of $J$
\begin{equation}
    J(d) = \int \frac{d^2q}{(2\pi)^2}\,J_{q}(d),
    \label{eq:J-mode-decomposition}
\end{equation}
where
\begin{align}
    J_{q}(d) &= \frac{1}{4q} \int dz\,dz'\, \rho_{q}(z)^* \rho_{q}(z')\left[e^{-q|d+z'-z|} +e^{-q|d-z'+z|} \right] \\
    &= \int dz\,dz'\, \rho_{q}(z)^* G_{q,d}(z-z')\rho_{q}(z') \ ,
    \label{eq:Jq-definition}
\end{align}
and we have used
\begin{equation}
    \int\frac{dk_z}{2\pi}\, \frac{e^{ik_z(z-z')}}{q^2+k_z^2} = \frac{e^{-q|z-z'|}}{2q} \ .
    \label{eq:partial-Coulomb}
\end{equation}
and conveniently defined
\begin{equation}
    G_{q,d}(z-z')= \frac{e^{-q|d+z'-z|} +e^{-q|d-z'+z|}}{4q} \ .
\end{equation}
which satisfies the following identity
\begin{equation}
     G_{q,d}(u+a)+ G_{q,d}(u-a) = G_{q,d+a}(u)+ G_{q,d-a}(u)
\end{equation}
and the Green's function identity
\begin{equation}
    \partial_z\partial_{z'} G_{q,0}(z-z') =\delta(z-z')-\frac q2e^{-q|z-z'|} \ .
\end{equation}

Let us now define
\begin{equation}
    F_{ q}(z)
=\int_{z-\Delta x/2}^{z+\Delta x/2}\rho_{ q}(s)\,ds  \ ,
\end{equation}
which naturally has support on $I=[-\Delta x/2,L+\Delta x/2]$, and satisfies $F_q(-\Delta x/2)=F_q(L+\Delta x/2)=0$. Then
\begin{equation}
    \partial_z F_q(z) = \rho_q(z+\Delta x/2)-\rho_q(z-\Delta x/2) \ .
\end{equation}

We now have all the ingredients to complete the proof. We call
\begin{equation}
     B_q(d)\equiv2J_q(d)-J_q(d+\Delta x)-J_q(d-\Delta x)\ .
\end{equation}
Then it is useful to integrate by parts
\begin{align}
    B_q(0)& \equiv2J_q(0)-J_q(\Delta x)-J_q(\Delta x)\\
    &= \int dz\,dz'\, \rho_{q}(z)^* \left[2G_{q,0}(z-z')-G_{q,\Delta x}(z-z')-G_{q,\Delta x}(z-z')\right]\rho_{q}(z') \\
    &= \int dz\,dz'\, F'_{q}(z)^* G_{q,0}(z-z') F'_{q}(z')\\
    &= \int dz\,dz'\, F_{q}(z)^* \partial_z\partial_{z'}G_{q,0}(z-z') F_{q}(z')\\
&=\int|F_{q}(z)|^2\,dz
-\frac q2\int\int
e^{-q|z-z'|}
F_{q}(z)^*F_{q}(z')\,dz\,dz',
\label{eq:local-increment-positive}
\end{align}

Adding and subtracting
\begin{equation}
    \frac{1}{2}\int dz \left(e^{-q(z+\Delta x/2)}+ e^{-q(L+\Delta x/2-z)}\right)|F(z)|^2 =\frac{I_+ + I_-}{2} \ , \qquad I_\pm = \int_I w_\pm(z) |F(z)|^2  dz \ ,
\end{equation}
where
\begin{equation}
    w_+(z)=e^{-q(L+\Delta x/2-z)} \ , \qquad w_-(z)=e^{-q(z+\Delta x/2)} \ ,
\end{equation}
allows one to rewrite the term to be bounded as a positive sum of squares
\begin{equation}
B_q(0)=
\frac q4\int \int e^{-q|z-z'|}
\left|F_{q}(z)-F_{q}(z')\right|^2\,dz\,dz'+\frac{1}{2}\int
\left[e^{-q(z+\Delta x/2)}+e^{-q(L+\Delta x/2-z)}\right]
|F_{q}(z)|^2\,dz \ .
\end{equation}
This gives the useful bound
\begin{equation}
    B_q(0) \geq \frac{I_-+I_+}{2}\geq\sqrt{I_+I_-}\geq 0 \ .
\end{equation}
Similarly, we can integrate by parts to obtain, under $d\geq L+\Delta x$
\begin{equation}
     B_q(d)=-\frac{q}{2}e^{-q(d-L-\Delta x)} \text{Re}\left[A^*_+ A_- \right] \ ,
\end{equation}
where we have defined
\begin{equation}
    A_\pm = \int w_\pm F_{q}(z) \,dz \ .
\end{equation}
This implies
\begin{equation}
    |B_{q}(d)|\leq \frac{q}{2}e^{-q(d-L-\Delta x)} |A_-| |A_+| 
\end{equation}

Finally, the weighted Cauchy-Schwarz inequality yields
\begin{equation}
    \left|A_\pm\right|^2
\le I_\pm\int w_\pm\,dz
=I_\pm\,\frac{1-e^{-q(L+\Delta x)}}q \ ,
\end{equation}
meaning that 
\begin{equation}
    |A_-| |A_+|\leq \frac{1-e^{-q(L+\Delta x)}}{q} \sqrt{I_+ I_-} \ .
\end{equation}
Combining everything, we get the following chain of inequalities
\begin{equation}
    |B_{q}(d)|\leq \frac{q}{2}e^{-q(d-L-\Delta x)} |A_-| |A_+| \leq\frac{ e^{-q(d-L-\Delta x)}-e^{-qd}}{2} \sqrt{I_+ I_-} \leq \frac{B_q(0)}{2} .
\end{equation}
This implies the no-entanglement condition, since
\begin{equation}
C_*(d,\Delta x) =\frac{1}{4(2\pi)^2}\int d^2q\,B_q(d) \ .
\end{equation}
Note that, although we have dropped it at times for notational convenience, the integrals involving $F_q(z)$ are always understood to run over the support $I$.

\subsubsection{Spherical Masses}
We now explicitly compute $\mathcal{G}$ for the case of compact spheres, a common choice for BMV setups. Let the masses be homogeneous spheres of radius $R$
\begin{equation}
    \rho_R(x)=\frac{3}{4\pi R^3}\Theta(R-|x|) \ .
\end{equation}
It is possible to find the $J$-integral explicitly for two identical spheres:
\begin{equation}
    J_R(d)=\frac{1}{4\pi R} F(d/R) \ ,
\end{equation}
where
\begin{equation}
F(u)=
\begin{cases}
\displaystyle
\frac{6}{5}-\frac{u^2}{2}+\frac{3u^3}{16}-\frac{u^5}{160},
&0\leq u\leq 2,\\
\frac{1}{u},&u\geq 2.
\end{cases}
\end{equation}

It is then straightforward to see that for $d-\Delta x \geq 2R$, i.e. if the mass-density distributions do not overlap in any branch, $\mathcal{G} = C_*(0,\Delta x)/|C_*(d,\Delta x)| \geq 4  \geq 2 $. 
Whenever the superposition is small, one finds to leading order that 
\begin{equation}
   \mathcal{G}  = \lim_{\Delta x \to 0 }  \frac{C_*(0,\Delta x)}{|C_*(d,\Delta x)|} = \frac{d ^3}{2 R^3} \geq 4
\end{equation}

Here, we do not claim that any two bodies, with no material overlap in any branch, do not get entangled if the DD trade-off is saturated. However, scenarios in which this could happen -- e.g. exploiting the nonlocality of the DP kernel -- would need to be highly engineered, as indicated by the sphere setup, which sits comfortably above the threshold.

\subsubsection{Gaussian Matter Distributions}
What happens if the object is not compact? Let us take the simplest computational example: normalized Gaussian profiles:
\begin{equation}
    \rho_{R_0}(x)
=\frac{e^{-| x|^2/(2R_0^2)}}{(2\pi R_0^2)^{3/2}} \ ,
\end{equation}
and direct integration gives
\begin{equation}
    J_{R_0}(d)=\frac{\mathrm{erf}(d/(2R_0))}{4\pi d} \ .
\end{equation}
The no-entanglement condition
\begin{equation}
    \left|2J_{R_0}(d)-J_{R_0}(d+\Delta x)-J_{R_0}(d-\Delta x)\right| \leq
\sqrt{2}\left[J_{R_0}(0)-J_{R_0}(\Delta x)\right] \ 
\end{equation}
cannot be explicitly solved for $d$. However, numerical solutions can be easily found. For example, in the small $\Delta x/R_0$ limit, the condition reduces to
\begin{equation}
    |J_{R_0}''(d)|\leq -\frac{J_{R_0}''(0)}{\sqrt{2}} \ ,
\end{equation}
where the leading $\Delta x$ contribution comes at fourth order. Solving numerically, we find that the leading no-entangling threshold at small superposition size is $d\gtrapprox 0.83 R_0$. In the opposite limit,

Although no BMV experiment is ever going to use mass densities with Gaussian profiles for the nanoparticles, this bound becomes relevant when the actual mass density of the particle enters the evolution law after smearing, as in the DP model, and the smearing length is much greater than the radius of the nanoparticle. The threshold we find is in quantitative agreement with the one found in Refs.~\cite{trillo_diosi_2025,entanglement_angeli_2025}.

\section{Model-independent bound on the entangling phase rate from smearings}
\label{app:phase_rate_bound}
In this appendix we bound the entangling phase rate relative to its bare Newtonian prediction. We consider two identical, compact, spherically symmetric matter distributions, with positive normalized profile $\rho$, and a positive, normalized, spherically symmetric smearing function $g$. The bare configurations are plane-separated,
$d-\Delta x\geq 2R$, with $0<\Delta x<d$, and the noise kernel obeys
the DD trade-off $D(k)\geq D_*(k)$.
Under these assumptions, we obtain the bound
\begin{equation}
  -0.137 < \frac{\dot\phi_e[g]}{\dot\phi_{e,*}} < 0
  \qquad\Longrightarrow\qquad
  -1.137 < \epsilon_\phi < -1
  \label{eq:phase-rate-final-bound}
\end{equation}
whenever entanglement is generated. Thus, the entangling phase has the opposite sign to the bare Newtonian prediction and is suppressed in magnitude by at least $86\%$. 

\subsection{Fourier-space entanglement condition}

 Let $\rho_g(k)=\rho(k)g(k)$. Positivity and normalization imply
\begin{equation}
  0\leq |\rho_g(k)|^2\leq 1.
\end{equation}
Define
\begin{align}
  L_{\Delta x}(k,d)
  &\equiv \mathrm{sinc}(kd) -\frac{1}{2}\mathrm{sinc}\left(k(d+\Delta x)\right)
 -\frac{1}{2}\mathrm{sinc}\left(k(d-\Delta x)\right) \ , \label{eq:L_Delta_def}
  \\
  L_{\Delta x}(k,0)
  &=1-\mathrm{sinc}(k\Delta x) \ .
\end{align}
Performing the angular integrals gives
\begin{align}
  C_g(d,\Delta x)
  &=\frac{1}{4\pi^2}\int_0^\infty dk\, |\rho_g(k)|^2 L_{\Delta x}(k,d),\\
  C_g(0,\Delta x)  &=\frac{1}{4\pi^2}\int_0^\infty dk\, |\rho_g(k)|^2 L_{\Delta x}(k,0).
\end{align}
Under the DD trade-off, entanglement requires $|C_g(d,\Delta x)|>C_g(0,\Delta x)/\sqrt{2}$. We also use the following inequality
\begin{equation}
  C_g(d,\Delta x)\geq-\frac{1}{2} C_g(0,\Delta x) \ ,
  \label{eq:negative_exclusion}
\end{equation}
which excludes the negative branch. Its proof is presented at the end of Appendix~\ref{app:force_bound}. Therefore, only the positive branch can entangle, and
\begin{equation}
  \int_0^\infty dk\,|\rho_g(k)|^2 \left[L_{\Delta x}(k,d) -\frac{1}{\sqrt{2}}L_{\Delta x}(k,0)\right]>0 \ .
  \label{eq:phaserate_entanglement}
\end{equation}

Bare plane separation and the shell theorem give
\begin{equation}
  C_*(d,\Delta x) =\frac{1}{16\pi}\left[  \frac{2}{d}-\frac{1}{d+\Delta x}-\frac{1}{d-\Delta x}\right] =-\frac{\Delta x^2}{8\pi d(d^2-\Delta x^2)}<0.
  \label{eq:phase-rate-bare-C}
\end{equation}
Since $\dot\phi_e[g]/\dot\phi_{e,*}=C_g(d,\Delta x)/C_*(d,\Delta x)$, the rates have opposite signs (since, as we have argued, $C_g\geq 0$), hence
\begin{equation}
    R_\phi\equiv
  -\frac{\dot\phi_e[g]}{\dot\phi_{e,*}}\geq 0 \ .
\end{equation}
Further, their positive magnitude ratio is
\begin{equation}
R_\phi =\frac{2d(d^2-\Delta x^2)}{\pi\Delta x^2}
  \int_0^\infty dk\,|\rho_g(k)|^2 L_{\Delta x}(k,d).
  \label{eq:phaserate_ratio}
\end{equation}

\subsection{A smearing-independent bound}

For any $\lambda>0$, we can add $\lambda$ times the positive integral in Eq.~\eqref{eq:phaserate_entanglement} to the integral in Eq.~\eqref{eq:phaserate_ratio} to find a useful upper bound. Indeed, using positivity and normalisation of the smearing function
\begin{align}
  \int_0^\infty dk\,|\rho_g(k)|^2 L_{\Delta x}(k,d) &<\int_0^\infty dk\,|\rho_g(k)|^2 \left[(1+\lambda)L_{\Delta x}(k,d) -\frac{\lambda}{\sqrt{2}}L_{\Delta x}(k,0)\right]\\
  &\leq\int_0^\infty dk\, \left[(1+\lambda)L_{\Delta x}(k,d) -\frac{\lambda}{\sqrt{2}}L_{\Delta x}(k,0)\right]_+.
\end{align}
Here $[z]_+=\max(z,0)$. Hence
\begin{equation}
R_\phi < B^{(\phi)}_\lambda(d,\Delta x) \equiv\frac{2d(d^2-\Delta x^2)}{\pi\Delta x^2} \int_0^\infty dk\, \left[(1+\lambda)L_{\Delta x}(k,d) -\frac{\lambda}{\sqrt{2}}L_{\Delta x}(k,0)\right]_+.
  \label{eq:phaserate_bound}
\end{equation}
This bound is independent of both the matter profile and the smearing function, and can be optimized over $\lambda$.

To find an explicit (although not necessarily optimal) bound, choose $\lambda=24$. Then
\begin{equation}
  \mathcal{F}_{d,\Delta x}(k) \equiv 25L_{\Delta x}(k,d) -12 \sqrt{2}L_{\Delta x}(k,0).
  \label{eq:phase-rate-F}
\end{equation}
Note that, as we prove in the last subsection of this appendix, for $kd>9$
\begin{equation}
L_{\Delta x}(k,d)\le\frac{2}{3}L_{\Delta x}(k,0) \ ,
\end{equation}
implying that in the same range 
\begin{equation}
    \mathcal{F}_{d,\Delta x}(k)\leq 0
     \label{eq:F_nonpositive}
\end{equation}

The angular representation of $L_{\Delta x}$ yields
\begin{equation}
  \frac{\mathcal{F}_{d,\Delta x}(k)}{\Delta x^2} =\frac{1}{k}\int_0^k dq\ \left(25\cos(qd)-12\sqrt{2}\right)  \frac{1-\cos(q\Delta x)}{\Delta x^2}.
  \label{eq:phase-rate-F-integral}
\end{equation}
First, for $0<\Delta x\leq d$ and $qd>0$
\begin{equation}
    0\leq \frac{1-\cos(q\Delta x)}{\Delta x^2}\leq \frac{q^2}{2}
\end{equation}
Moreover, for the interval $0\leq qd\leq 2\pi$
\begin{equation}
    \frac{1-\cos(q\Delta x)}{\Delta x^2} \geq \frac{1-\cos(q d)}{d^2} \ ,
\end{equation}
from the monotonicity of $\mathrm{sinc}(z)$ on $0\leq z\leq\pi$. Then:
\begin{equation}
  \frac{1-\cos(qd)}{d^2} \leq\frac{1-\cos(q\Delta x)}{\Delta x^2}  \leq\frac{q^2}{2}  \ ,
  \label{eq:phase-rate-cosine-envelope}
\end{equation}
and consequently
\begin{equation}
  \frac{\mathcal{F}_{d,\Delta x}(k)}{\Delta x^2} \leq \mathcal{H}_d(k),
  \label{eq:F_bound_H}
\end{equation}
where we have defined
\begin{equation}
  \mathcal{H}_d(k)\equiv\frac{1}{kd^3}\int_0^{kd} dz\ M(z).
  \label{eq:H_bound}
\end{equation}
and
\begin{equation}
\begin{aligned}
M_\phi(z)=& [25\cos(z)-12\sqrt{2}]_+\frac{z^2}{2}\\
&- [12\sqrt{2}-25\cos(z)]_+ \left[1-\cos(z)\right]\,
\Theta(2\pi-z),
\end{aligned}
\end{equation}

Combining Eqs.~\eqref{eq:phaserate_bound}, \eqref{eq:F_bound_H}, and \eqref{eq:F_nonpositive} gives
\begin{align}
R_\phi < \frac{2d(d^2-\Delta x^2)}{\pi} \int_0^{9/d}[\mathcal H_d(k)]_+\,dk \leq \frac{2d^3}{\pi} \int_0^{9/d}[\mathcal H_d(k)]_+\,dk.
  \label{eq:phaserate_bound_gen}
\end{align}
where in the last step we rounded conservatively the numerical estimation for the bound. Note that the bound is manifestly $d$-independent since
\begin{equation}
    \frac{2d^3}{\pi} \int_0^{9/d}[\mathcal H_d(k)]_+\,dk = \frac{2}{\pi} \int_0^{9}\left[\int_0^u M_\phi(z) dz\right]_+\,\frac{du}{u} 
\end{equation}

Defining the bound on $\epsilon_\phi$ for convenience instead
\begin{equation}
    B^{(\phi)}  =-1-\frac{2}{\pi} \int_0^{9}\left[\int_0^u M_\phi(z) dz\right]_+\,\frac{du}{u} \approx -1.137\ ,
    \label{eq:B_phi}
\end{equation}
where in the last step we have reported the result of numerical quadrature of the integral. Combining this with the definition of $\epsilon_\phi=-R_\phi-1$ yields the claimed
\begin{equation}
    -1.137\approx B^{(\phi)}\leq \epsilon_\phi \leq -1 \ .
\end{equation}

\subsection{Proving $\mathcal{F}_{d,\Delta x}(k)\leq 0$ for $kd>9$}
For $\lambda=24$, we have
\begin{equation}
    \mathcal{F}_{d,\Delta x}(k)=25L_{\Delta x}(k,d)-12\sqrt{2}\,L_{\Delta x}(k,0).
\end{equation}
We assume $0<\Delta x\leq d$ (the plane-separation condition). Since $L_{\Delta x}(k,0)\geq0$ and $50/3-12\sqrt{2}<0$, it suffices to establish
\begin{equation}
    L_{\Delta x}(k,d)\leq\frac{2}{3}L_{\Delta x}(k,0).
    \label{eq:tail_ratio}
\end{equation}

First, suppose $0<k\Delta x<\pi$. Integrating by parts the angular representation of $L$ gives
\begin{align}
    L_{\Delta x}(k,d)&=\int_0^1du\,\cos(kdu)\left[1-\cos(k\Delta x\,u)\right]\\
    &=\frac{\sin(kd)}{kd}\left[1-\cos(k\Delta x)\right]-\frac{\Delta x}{d}\int_0^1du\,\sin(kdu)\sin(k\Delta x\,u).
\end{align}
In this range $\sin(k\Delta x\,u)\geq0$. Using $\sin(kd)\leq1$ and $\sin(kdu)\geq -1$, we obtain
\begin{align}
    L_{\Delta x}(k,d)&\leq\frac{1-\cos(k\Delta x)}{kd}+\frac{\Delta x}{d}\int_0^1du\,\sin(k\Delta x\,u)\\
    &=\frac{2[1-\cos(k\Delta x)]}{kd}.
\end{align}
Moreover, concavity of the sine implies $\sin(k\Delta x\,u/2)\geq u\sin(k\Delta x/2)$ for $0\leq u\leq1$, and therefore
\begin{equation}
    L_{\Delta x}(k,0)=\int_0^1du\,\left[1-\cos(k\Delta x\,u)\right]\geq[1-\cos(k\Delta x)]\int_0^1du\,u^2=\frac{1-\cos(k\Delta x)}3.
\end{equation}
Combining these inequalities yields
\begin{equation}
    L_{\Delta x}(k,d)\leq\frac6{kd}L_{\Delta x}(k,0)\leq\frac23L_{\Delta x}(k,0),\qquad kd\geq9.
\end{equation}

Next, consider $k\Delta x\geq\pi$ instead. The elementary bounds
\begin{equation}
    |\mathrm{sinc}z|\leq\frac{1}z\quad(z>0),\qquad \mathrm{sinc}z\geq-\frac{1}\pi\quad(z\geq0)
\end{equation}
give
\begin{align}
    L_{\Delta x}(k,d)&\leq\frac{1}{kd}+\frac{1}{2k(d+\Delta x)}+\frac{1}{2\pi}\leq\frac3{2kd}+\frac{1}{2\pi},\\
    L_{\Delta x}(k,0)&\geq1-\frac{1}{k\Delta x}\geq 1-\frac{1}\pi.
\end{align}
Hence, for $kd\geq 9$,
\begin{equation}
    L_{\Delta x}(k,d)\leq\frac{1}{6}+\frac{1}{2\pi}<\frac{2}{3}\left(1-\frac{1}\pi\right)\leq\frac{2}{3}L_{\Delta x}(k,0),
\end{equation}
where the strict inequality follows trivially from $\pi>7/3$. This establishes Eq.~\eqref{eq:tail_ratio} in both cases, and consequently
\begin{equation}
    \mathcal{F}_{d,\Delta x}(k)\leq\left(\frac{50}{3}-12\sqrt{2}\right)L_{\Delta x}(k,0)\leq0,\qquad kd\geq9.
\end{equation}

\section{Model-independent force bound}
\label{app:force_bound}
In this appendix we prove that, for any positive, normalised and spherically symmetric smearing function $g$, GME in classical theories of gravity is always accompanied by a suppression of at least 84\% from the Newtonian inverse-square law for the gravitational force between spherically symmetric masses. 

To prove this, let us first define
\begin{equation}
    V(d)=\int\frac{h(z)}{|d-z|}\,d^3z = 4\pi J_g(d) , \qquad h=\rho*\rho*g*g
\end{equation}
such that we can rewrite the gravitational potential energy of the smeared spherical masses as
\begin{equation}
    U_g(d) =- GM^2 V(d) \ .
\end{equation} 
Note that $J_g$ is the $J$ in Equation~\ref{eq:J-definition} evaluated with the smeared mass-density $\rho_g$ instead. The equation matches the gravitational potential energy between a spherically symmetric object with normalised density $h$ and total mass $M$, and a point particle of the same mass a distance $d$ away from its center. Therefore, since $h(z)$ is spherically symmetric, the force between the two particles corresponds to the “mass” $Q(d)$ of $h$ contained within the sphere of radius $d$, i.e.
\begin{equation}
    F_g(d)=\frac{GM^2Q(d)}{d^2} \ ,
\end{equation}
meaning that it is indeed
\begin{equation}
   1-\delta_F(d)\equiv  \frac{F_g(d)}{F_N(d)} =4\pi\int_0^d r^2  h(r) dr=Q(d)
\end{equation}
that controls the ratio between the Newtonian and the smeared forces.

The entanglement condition amounts to 
\begin{equation}
   | 2V(d)-V(d+\Delta x)-V(|d-\Delta x|)| \geq \sqrt{2}(V(0)-V(\Delta x) )\ ,
\end{equation}
but, as we prove in the last subsection of this appendix, the negative branch can never be satisfied. Hence, we can remove the absolute value, meaning
\begin{equation}
     2V(d)-V(d+\Delta x)-V(|d-\Delta x|)\geq \sqrt{2}(V(0)-V(\Delta x))
\end{equation}
is necessary for the two qubits to entangle under the DD trade-off. However, we can now show that this in turn implies an upper bound for $Q(d)$ in the range of $d$ for which entanglement is generated.

To do this, it is useful to go into momentum space (all the angular integrals will be performed, since we are using the simplifying assumption of spherical symmetry for $\rho(k)$, on top of the physical requirement of spherical symmetry for the smearing functions). Then, it is straightforward to see
\begin{align}
    2V(d)-V(d+\Delta x)-V(d-\Delta x) &= \frac{4}{\pi}\int_0^\infty dk |\rho(k)g(k)|^2 \left[\mathrm{sinc}(kd) -\frac{1}{2}\mathrm{sinc}( k(d+\Delta x))-\frac{1}{2}\mathrm{sinc}\left( k(d-\Delta x)\right) \right]\\
    & =\frac{4}{\pi}\int_0^\infty dk |\rho(k)g(k)|^2 L_{\Delta x}(k,d) \ ,
\end{align}
where in the last line we have identified the same function $L_{\Delta x}(k,d)$ first defined in Equation~\ref{eq:L_Delta_def}. Consequently, the entanglement condition in Fourier space is
\begin{equation}
    \int_0^\infty dk |\rho(k)g(k)|^2 \left[L_{\Delta x}(k,d)-\frac{1}{\sqrt{2}}L_{\Delta x}(k,0) \right] \geq 0 \ .
\end{equation}
This is conveniently close to the momentum-space representation of the force ratio:
\begin{align}
    Q(d) &=\frac{2d}{\pi}\int_0^\infty dk |\rho(k)g(k)|^2 \left[\mathrm{sinc}(kd)-\cos(kd) \right]\\
    &\leq \frac{2d}{\pi}\int_0^\infty dk |\rho(k)g(k)|^2 \left[\mathrm{sinc}(kd)-\cos(kd) +\lambda \left(L_{\Delta x}(k,d)-\frac{1}{\sqrt{2}}L_{\Delta x}(k,0) \right)\right]
\end{align}
for any $\lambda\geq 0$ since, in the entanglement regime, the term we added in the last line is always positive. Moreover, since $0\leq|\rho(k) g(k)|^2 \leq1$, we can conclude
\begin{equation}
    Q(d) \leq B^{(F)}_{\lambda}(d,\Delta x) \equiv \frac{2d}{\pi}\int_0^\infty dk \left[\mathrm{sinc}(kd)-\cos(kd) +\lambda \left(L_{\Delta x}(k,d)-\frac{1}{\sqrt{2}}L_{\Delta x}(k,0) \right)\right]_+ \ \ ,
\end{equation}
where $f_+=\max(f,0)$.
This bound is a function of both $\lambda$ \emph{and} $\Delta x$. Any valid choice of $\lambda>\sqrt{2}$ (to guarantee convergence of the tail for any $\Delta x$) would give a bound, but the optimal one comes from optimising over $\lambda$, which is an easy numerical task. 

\subsection{Universal bound}
This tells us that if entanglement is measured for a $(d,\Delta x)$ pair, then there is a consequent deviation in the average force felt by two classical masses at distance $d$. Doing the optimisation is in general hard, but we now derive a universal bound for the force deviation at $d$  given a positive entanglement result for finite $\Delta x$ in the plane-separated geometry. The bound, although sub-optimal, is still significant. In the next subsection we find the optimal bound if entanglement is measured for small $\Delta x$.

For now, consider the choice $\lambda = 12 d^2/\Delta x^2$, and conveniently rewrite
\begin{equation}
    \frac{12 d^2}{\Delta x^2}\left(L_{\Delta x}(k,d)-\frac{1}{\sqrt{2}}L_{\Delta x}(k,0) \right) = \frac{2}{kd}\int _0^{kd} dz \left(\cos z-\frac{1}{\sqrt{2}}\right) F_{\Delta x/d}(z) \ ,
\end{equation}
where
\begin{equation}
    F_{\Delta x/d}(z) =\frac{6d^2}{\Delta x^2}\left(1-\cos(z\Delta x/d)\right)
=3z^2\mathrm{sinc}^2(z\Delta x/2d) \ .
\end{equation}
This rewriting helps us to find a simple integrable function that bounds $Q(d)$ in the plane-separated geometry ($0\leq\Delta x/d\leq 1$), independent of the superposition size. Indeed, in that range
\begin{equation}
    0\leq F_{\Delta x/d}(z)\leq 3z^2 \ .
\end{equation}
Moreover
\begin{equation}
    F_{\Delta x/d} \geq 6(1-\cos z) \, \qquad 0\leq z\leq 2\pi \ .
\end{equation}
This allows us to write
\begin{equation}
    \left(\cos z-\frac{1}{\sqrt{2}}\right)F_{\Delta x/d}(z) \leq M(z)\ ,
\end{equation}
where
\begin{equation}
    M_F(z)= 3z^2\left(\cos z-\frac{1}{\sqrt{2}}\right)_+
-6(1-\cos z)\left(\frac{1}{\sqrt{2}}-\cos z\right)_+\,
\Theta(2\pi-z) \ ,
\end{equation}
The bound is now $\Delta x$-independent -- certifying entanglement at \emph{any} $\Delta x$ implies the following bound on the forces between two classical particles at the same average distance $d$:
\begin{equation}
    B^{(F)}\equiv B^{(F)}_{12d^2/\Delta x^2} \leq \frac{2d}{\pi}\int_0^{12/d} dk \left[\mathrm{sinc}(kd)-\cos(kd) + \dfrac{2}{kd}\int_0^{kd} M(z) dz\right]_+ \ .
\end{equation}
since one can now show that
\begin{equation}
\label{eq:tail_bound}
   \mathrm{sinc}(kd)-\cos(kd)+ \frac{12 d^2}{\Delta x^2}\left(L_{\Delta x}(k,d)-\frac{1}{\sqrt{2}}L_{\Delta x}(k,0) \right) <0 \ ,\qquad kd>12 \ .
\end{equation}
We prove this key inequality in two steps. First, take $0\leq k\Delta x<\pi$. Then, integrating by parts and using $\sin(k\Delta x)\geq 0$ in this range, we get
\begin{equation}
    L_{\Delta x}(k,d) = \int_0^1 du \cos (kd \ u)(1-\cos(k\Delta x \ u) )\leq \frac{2(1-\cos(k\Delta x))}{kd} \ .
\end{equation}
Similarly, from concavity of the sine,
\begin{equation}
    L_{\Delta x}(k,0) = \int_0^1 du (1-\cos(k\Delta x \ u) \geq \int_0^1 du u^2 (1-\cos(k\Delta x))=\frac{ (1-\cos(k\Delta x))}{3} \ ,
\end{equation}
implying
\begin{equation}
    L_{\Delta x}(k,d) \leq\frac{6}{kd} L_{\Delta x}(k,0) \leq  \frac{1}{2}L_{\Delta x}(k,0)
\end{equation}
in the range, for $kd>12$. Moreover, using $\sin(ax/2) \geq ax/\pi$, we see
\begin{equation}
    L_{\Delta x}(k,0) \geq \frac{2\Delta x^2}{3d^2} \ .
\end{equation}
We can show the same bounds for $k\Delta x\geq\pi$, using 
\begin{equation}
    \mathrm{sinc}(x)\geq -\frac{1}{\pi} \ , \qquad \mathrm{sinc}(x)\leq \frac{1}{\pi} \ \textrm{if} \ x\geq\pi .
\end{equation}
Combining these inequalities gives the desired Eq.~\ref{eq:tail_bound}. 

Hence, we can simply use numerical quadrature to evaluate the integral on the RHS, obtaining
\begin{equation}
    Q(d)\leq 0.1554 \lessapprox0.16 \ ,
\end{equation}
rounding up to account for numerical precision.

\subsection{Small $\Delta x$ bound}
If entanglement is measured in the $\Delta x \to 0$ limit, the optimal bound can in fact be derived exactly. In this limit
\begin{equation}
    L_{\Delta x}(k,d)=\frac{\Delta x^2 k^2}{6} b(kd) \ , \qquad b(x) \equiv 3\frac{(x^2-2) \sin(x) + 2 x \cos( x)}{x^3}
\end{equation}
with $b(0)=1$. Then, the bound in the $\Delta x \to 0$ limit becomes
\begin{equation}
    B_{\mu}(d) = \frac{2d}{\pi}\int_0^\infty dk \max\left[\mathrm{sinc}(kd)-\cos(kd) +\mu k^2\left(b(kd)-\frac{1}{\sqrt{2}}\right),0\right] 
\end{equation}
where we conveniently rescaled $\mu=\lambda \Delta x^2/6$.

To find the optimal $\mu_\star$, first assume that the integrand is non-zero only within a single region $A=(0,k_*)$ for the optimal choice of $\mu=\mu_*$. This implies
\begin{equation}
    B_{\mu}(d) = \frac{2d}{\pi}\int_0^{k_*} dk \left[\mathrm{sinc}(kd)-\cos(kd) +\mu_* k^2\left(b(kd)-\frac{1}{\sqrt{2}}\right)\right] \ .
\end{equation}
Then, we can differentiate to find $\mu_*$ 
\begin{equation}
   \mu_* = -\frac{\mathrm{sinc}(k_*d)-\cos(k_*d)}{k_*^2\left(b(k_*d)-\frac{1}{\sqrt{2}}\right)} 
\end{equation}
where $k_*$ satisfies
\begin{equation}
    \int_0^{k_*} dk k^2\left[b(k)-\frac{1}{\sqrt{2}}\right]=\
-3k_*\cos(k_*)+9\sin (k_*)
-6\mathrm{Si}(k_*)-\frac{1 k_*^3}{3\sqrt{2}}=0 \ .
\end{equation}
Solving numerically, one finds $k_*d\approx1.32$,
and, consequently, $\mu_*/d^2\approx 1.54$. Plugging this back into the integrand, it is straightforward to see that there is indeed a single zero for this choice of $\mu$, making the initial assumption valid. We can then plug this back into the expression for the bound, finding
\begin{equation}
    \lim_{\Delta x \to 0}Q(d) \lessapprox 0.15 \ ,
\end{equation}
implying at least an 85\% reduction in the observed Newtonian force between two equal, spherically symmetric masses that can be entangled in a theory of classical gravity for small superpositions.

\subsection{Excluding the negative branch, $ C_g(d,\Delta x)\geq-\frac{1}{2} C_g(0,\Delta x)$}
Here we prove the key inequality of Eq.~\ref{eq:negative_exclusion}, or, equivalently
\begin{equation}
    2V(d)-V(d+\Delta x)-V(|d-\Delta x|) \geq - [V(0)-V(\Delta x)] \ .
\end{equation}

First, define the marginal of the smeared mass density along the superposition axis
\begin{equation}
    p(z) = 2\pi \int_{|z|}^\infty r \ (g*\rho)(r) dr \ ,
\end{equation}
which, since $g\geq 0$ and $\rho\geq 0$, is even, non-negative and non-increasing for $z\geq0$. Then, define
\begin{equation}
    \psi_{\Delta x}(z) = p(z-\Delta x/2)-p(z+\Delta x/2) \ .
\end{equation}
To connect with the proof on plane-separated compact masses, $p(z)$ is equivalent to $\rho_0(z)$, after smearing. Then,
\begin{equation}
    H(s) = \int p(z)p(z+s) dz = 2\pi\int_{|s|}^\infty r h(r) dr \ ,
\end{equation}
meaning, by the shell theorem, that
\begin{equation}
    V(r)= \frac{2}{r}\int_0^r H(s) ds = 2 \int_0^1 H(t r) dt \ .
\end{equation}
Then, in terms of $H$, we can rewrite
\begin{equation}
\begin{split}
    2V(d)-V(d+\Delta x)-V(d-\Delta x) &= 2\int_0^1 ds\left(2 H(ds)- H(ds+s\Delta x)- H(ds-s\Delta x)\right) \\
    & = 2\int_0^1 ds \int_{-\infty}^\infty dz  \ \psi_{s\Delta x}(z) \psi_{s\Delta x}(z+sd) \ ,
\end{split}
\end{equation}
and similarly
\begin{equation}
    2\left[V(0) -V(\Delta x)\right] = 4 \int_0 ^1 ds \left(H(0)- H(s\Delta x)\right)) = 2 \int_0^1 ds \int_{-\infty}^{\infty} dz  \ \psi_{s \Delta x}(z)^2 \ .
\end{equation}
Since $\psi_a$ is odd and nonnegative for positive arguments, the integrand is also nonnegative outside $0\leq z\leq sd$. Then, it is straightforward to prove with Cauchy--Schwarz that
\begin{equation}
\begin{split}
    \int_{-\infty}^\infty dz  \ \psi_{s\Delta x}(z) \psi_{s\Delta x}(z-sd) &\geq  \int_{0}^{sd} dz  \ \psi_{s\Delta x}(z) \psi_{s\Delta x}(z-sd) \\
    & \geq - \frac{1}{2}\int_{0}^{sd} dz  \left[\psi_{s\Delta x}(z)^2 +\psi_{s\Delta x}(z-sd) ^2\right] \\
    &=- \int_{0}^{sd} dz  \ \psi_{s\Delta x}(z) ^2 \\
    &\geq- \frac{1}{2}\int_{-\infty}^{\infty} dz  \ \psi_{s\Delta x}(z) ^2 \ ,
\end{split}
\end{equation}
which then implies (the autocorrelation is invariant when replacing $z-sd$ with $z+sd$)
\begin{equation}
\begin{split}
    2V(d)-V(d+\Delta x)-V(d-\Delta x) &= 2\int_0^1 ds \int_{-\infty}^\infty dz  \ \psi_{s\Delta x}(z) \psi_{s\Delta x}(z+sd) \\
    &\geq - \int_0^1 ds \int_{-\infty}^{\infty} dz  \ \psi_{s \Delta x}(z)^2 \ = -\left[V(0) -V(\Delta x)\right] 
\end{split}
\end{equation}
concluding the proof.

\section{Bounding $R_0$ from precision experiments}
\label{app:R0_bound}

In this appendix, we first derive a bound on $R_0$ for smearing functions with finite fourth moments, given agreement with Newton's law up to a confidence bound $\delta_F$. Then, we show that for the specific choice of Bessel smearing, we can recover the Yukawa correction to the $1/r$ potential studied by~\cite{new_lee_2020}.
\subsection{A bound on $R_0$ from short-range experiments}
The fractional deviation from the Newtonian force between point particles for a chosen smearing function $g$ is
\begin{equation}
   1-\frac{F_g(d)}{F_N(d)} =\int_{|x|>d} h(x) \ d^3x  \ ,
\end{equation}
where $h=g*g$. Moreover, let
\begin{equation}
    \int d^3 x \ h(x) x^4 \leq (2c^4+30)R_0^4 \ ,
\end{equation}
and
\begin{equation}
\int d^3x h(x)x^2\ d^3x= 6R_0^2.
\end{equation}
We can split the integral above across the $|x|=d$ radius, and use Cauchy-Schwartz to show
\begin{equation}
6R_0^2\le d^2+R_0^2\sqrt{(2c^4+30)\delta}.
\label{eq:R0_bound_point}
\end{equation}
Rearranging, we find a universal conservative bound for the chosen family of bounded-kurtosis smearing kernels for an experimental observation of $\delta_F(d)\leq \delta$
\begin{equation}
    R_0 \leq \frac{d}{\sqrt{6-\sqrt{(2c^4+30)\delta}}} \approx\frac{d}{\sqrt6}\left[1+\frac{\sqrt{(2c^4+30)\,\delta}}{12}\right]  \ .
\end{equation}
Note that the bound can be inferred only for $c<\left(\frac{18}{\delta}-15\right)^{1/4} \ ,$ for a given sensitivity $\delta$. The leading term in the small-sensitivity expansion is independent of the actual value of $c$ and $\delta$
\begin{equation}
    R_0 \lessapprox 0.4 d \ .
\end{equation}
Moving away from the point-particle limit is straightforward (just repeat the same derivation with $g\to \rho*g$ instead), and an equivalent closed-form version of Equation~\ref{eq:R0_bound_point} can be found for the case in which the masses are spherical. However, the expression in cumbersome, and the gain one obtains is minimal, so we omit it here. Furthermore, if one is interested only in a conservative bound (although not optimal), the point-mass approximation provides it, since including the shape of the body would increase the predicted force deficit, at fixed smearing, hence implying a refined upper-bound from the same experiment would be stronger. 

\subsection{The Bessel smearing gives a Yukawa modification}
\label{app:yukawa}
Consider the normalized, spherically symmetric smearing kernel
\begin{equation}
g(r)=\frac{K_1(r/\lambda)}{2\pi^2\lambda^2 r}.
\label{eq:app_K1}
\end{equation}
Its Fourier transform is
\begin{equation} 
g(k)=\frac{1}{\sqrt{1+\lambda^2 k^2}}.
\end{equation}
Smearing both masses simply multiplies the Fourier transform of the Newtonian kernel by $ (k)^2$. The resulting expression, then, splits into two terms,
\begin{equation}
V_g(k)=\frac{4\pi}{k^2}\, g(k)^2=\frac{4\pi}{k^2}-\frac{4\pi}{k^2+\lambda^{-2}}.
\end{equation}
The first term is just the standard Newtonian kernel, whilst the second gives the Yukawa correction. Transforming back to position space, we obtain the pair potential
\begin{equation}
U_g(r)=-GM_1M_2 V_g(r)=-\frac{GM_1M_2}{r}\left(1-e^{-r/\lambda}\right).
\end{equation}
Thus the kernel in Eq.~\eqref{eq:app_K1} gives Eq.~\eqref{eq:yukawa_pot} with $\alpha=-1$.

One can also see why the $\alpha=-1$ value is singled out within the Yukawa family by the mass-smearing construction. For a general $\alpha$, matching the two potentials in Fourier space requires
\begin{equation}
 g(k)^2=1+\alpha\,\frac{\lambda^2 k^2}{1+\lambda^2 k^2}
\xrightarrow{k\to\infty}1+\alpha.
\end{equation}
For any integrable smearing density, $g(k)$ tends to zero at large $k$, so $\alpha=-1$ is the only possible amplitude. Values $-1<\alpha<0$ would require a point mass at the center of the smearing distribution, carrying a compensating fraction $\sqrt{1+\alpha}$ of the total mass.

Differentiating the pair potential gives the force ratio used in Appendix~\ref{app:force_bound},
\begin{equation}
\frac{F_g(r)}{F_N(r)}=1-\left(1+\frac{r}{\lambda}\right)e^{-r/\lambda}.
\end{equation}
The small-$k$ expansion of $g(k)$ then gives the radial moments $\langle r^2\rangle_g=3\lambda^2$ and $\langle r^4\rangle_g=45\lambda^4$, so $R_0=\lambda$ and $c^4=45$.

\section{Asymmetric Matter Distributions, Geometric Factors, and Decoherence Rates}
\label{app:unequal_masses}
 
Often, in experimental settings, the matter distribution that sources the gravitational field is different from the one that is used to sense it. For this reason, here we present the small adjustments needed for two different matter distributions.
 
Let $m_i(x,\hat{X}_i)=M_i \rho_i( x- \hat{X}_i)$, with $\int d^3x\,\rho_i(x)=1$ and $\hat{X}_i$ the centre of mass, be the two matter distributions each prepared in a superposition of (not necessarily equal) size $\Delta x_i$, with $i \in \{ 1,2\}$.
We place both bodies on the $z$ axis with body 1 above body 2, i.e.\ the branch positions are $X_1=(d\pm\frac{\Delta x_1}{2})\hat{z}$ and $X_2=\pm\frac{\Delta x_2}{2}\hat{z}$ (for the opposite placement replace $e^{-ik_zd}\to e^{+ik_zd}$, i.e.\ $J_{12}(r)\to J_{12}(-r)$ below, which is immaterial if each $\rho_i$ is symmetric under $z\to-z$).
Then, one finds the analogous non-entanglement condition
\begin{equation}
    \lambda^{\text{pt}}_- (t) \geq 0 \; \forall \; t \iff 4\Gamma_l^{(1)} \Gamma_l^{(2)}  -\Gamma_e^2 \geq \dot{\phi}_e^2
\end{equation}
where
\begin{align}
\dot{\phi}_e &= -\frac{4 M_1 M_2 }{\kappa}\int \frac{d^3k}{(2\pi)^3}\,\frac{\text{Re} \left[ \rho_1(k) \rho^*_2(k) e^{-i k_z d} \right]}{k^2}  \sin \left(\frac{k_z \Delta x_1}{2} \right) \sin \left(\frac{k_z \Delta x_2}{2} \right)
   \\
 \Gamma_l^{(i)}&=2 M_i^2 \int \frac{d^3k}{(2\pi)^3}\,
 |\rho_i(k)|^2 D(k) \sin^2 \left(\frac{k_z \Delta x_i}{2} \right)   \\
\Gamma_e&= 4 M_1 M_2 \int \frac{d^3k}{(2\pi)^3}\,
\text{Re} \left[ \rho_1(k) \rho^*_2(k) e^{-i k_z d} \right] D(k)  \sin \left(\frac{k_z \Delta x_1}{2} \right) \sin \left(\frac{k_z \Delta x_2}{2} \right) \label{eq:rate_cor_asym}
 \end{align}
Using $\Gamma_{l,*}^{(i)}=2M_i^2C^{ii}_*/\kappa$, $\Gamma_{e,*}=4M_1M_2C^{12}_*/\kappa$ and $\dot{\phi}_e=-4M_1M_2C^{12}_*/\kappa$, one finds
 \begin{equation}
     \chi_{\text{GME}} \equiv \frac{4\Gamma_l^{(1)} \Gamma_l^{(2)}  -\Gamma_e^2}{\dot{\phi}_e^2 } = \frac{4\Gamma_l^{(1)} \Gamma_l^{(2)}  -\Gamma_e^2}{4\Gamma_{l,*}^{(1)} \Gamma_{l,*}^{(2)}  -\Gamma_{e,*}^2 }
      \left[ \frac{C^{11}_*(0, \Delta x_1) C^{22}_*(0, \Delta x_2)}{ C^{12}_*(d, \Delta x_1,\Delta x_2)^2}-1 \right] \equiv \nu_{\text{exp}} \left[ \mathcal{G}^2 - 1\right]
 \end{equation}
 where $\nu_{\text{exp}}$ is the first ratio and $\mathcal{G}\equiv\sqrt{C^{11}_*(0,\Delta x_1)\,C^{22}_*(0,\Delta x_2)}\,/\,|C^{12}_*(d,\Delta x_1,\Delta x_2)|\geq1$ by Cauchy--Schwarz. For identical bodies, $\chi_{\text{GME}}$, $\nu_{\text{exp}}$ and $\mathcal{G}$ reduce to the definitions of the main text, Eq.~\eqref{eq:geo_cond}.
 The geometric factors are
 \begin{align}
     C^{ii}_*(0, \Delta x_i) &= \int \frac{d^3k}{(2\pi)^3}\,
 \frac{|\rho_i(k)|^2}{k^2} \sin^2 \left(\frac{k_z \Delta x_i}{2} \right) = \frac{1}{2} \left[ J_{ii}(0)-J_{ii}(\Delta x_i)\right]\\
     C^{12}_*(d, \Delta x_1, \Delta x_2) &= \int \frac{d^3k}{(2\pi)^3}\,
 \frac{\text{Re} \left[ \rho_1(k) \rho^*_2(k) e^{-i k_z d} \right]}{k^2}  \sin \left(\frac{k_z \Delta x_1}{2} \right) \sin \left(\frac{k_z \Delta x_2}{2} \right) \\
 &= \frac{1}{4} \left[J_{12} \left(d + \frac{\Delta x_1- \Delta x_2}{2} \right)+J_{12} \left(d - \frac{\Delta x_1- \Delta x_2}{2} \right)- J_{12} \left(d + \frac{\Delta x_1+ \Delta x_2}{2}\right)  - J_{12} \left(d - \frac{\Delta x_1+ \Delta x_2}{2} \right)   \right] \nonumber
 \end{align}
where we defined
\begin{align}
    J_{jj}(r) &= \int \frac{d^3k}{(2\pi)^3}\,\frac{|\rho_{j}(k)|^2}{k^2} \cos(k_z r) = \frac{1}{4\pi}\int d^3 x \int d^3 y\, \frac{\rho_j(x) \rho_j(y)}{|x - y + r\hat{z}|}   \\
     J_{12}(r) &= \int \frac{d^3k}{(2\pi)^3}\,\frac{\text{Re} \left[ \rho_1(k) \rho^*_2(k) {\rm e}^{- i k_z r} \right] } {k^2} = \frac{1}{4\pi}\int d^3 x \int d^3 y\, \frac{\rho_1(x) \rho_2(y)}{|x - y + r\hat{z}|}
\end{align}

For two homogeneous spheres of radii $R_1$ and $R_2$, the shell theorem gives $J_{12}(r) = 1/(4 \pi |r|)$ whenever $|r|\geq R_1+R_2$. If this holds in every branch, i.e.\ for $d-\frac{\Delta x_1+\Delta x_2}{2}\geq R_1+R_2$, then
\begin{equation}
    C^{12}_*(d, \Delta x_1, \Delta x_2) = -\frac{d \Delta x_1 \Delta x_2}{8\pi \left(d^2 - \frac{(\Delta x_1 + \Delta x_2 )^2}{4}\right) \left(d^2 - \frac{(\Delta x_1 - \Delta x_2 )^2}{4} \right) }
\end{equation}
For the self-energy, the two cases read
\begin{align}
    \Delta x_i \geq 2R_i &\implies C^{ii}_*(0, \Delta x_i) = \frac{1}{4 \pi} \left( \frac{3}{5 R_i}-\frac{1}{2 \Delta x_i} \right) \\
        \Delta x_i \leq 2R_i &\implies C^{ii}_*(0, \Delta x_i) =\frac{1}{4 \pi} \left(\frac{\Delta x_i^5}{320 R_i^6}-\frac{3 \Delta x_i^3}{32 R_i^4}+\frac{\Delta x_i^2}{4 R_i^3}\right)
\end{align}
When considering experiments three cases are regarding the distinguishable or not: the superposition is present in both the masses, in only one, or in none of them. In each case, by using the two possible $C^{jj}_*$, one can compute $\mathcal{G}$, and then, when classical, take the limit $\Delta x_i \to 0$.
We label a body ``q'' if its branches do not overlap, $\Delta x_i\geq 2R_i$ (first line of $C^{ii}_*$), and ``c'' if $\Delta x_i\to0$, the limit being taken on $\mathcal{G}^2$ using the second line of $C^{ii}_*$; in $\mathcal{G}^2_{cq}$ body 1 is superposed and body 2 is classical. All three cases assume $d-\frac{\Delta x_1+\Delta x_2}{2}\geq R_1+R_2$. In this form, one finds
\begin{align}
    \mathcal{G}_{qq}^2 &=\frac{4\left( \frac{3}{5 R_1}-\frac{1}{2 \Delta x_1} \right) \left( \frac{3}{5 R_2}-\frac{1}{2 \Delta x_2} \right)}{ d^2 \Delta x_1^2 \Delta x_2^2 }  \left(d^2 - \frac{(\Delta x_1 + \Delta x_2 )^2}{4}\right)^2 \left(d^2 - \frac{(\Delta x_1 - \Delta x_2 )^2}{4} \right) ^2 \\
    &= \left(\frac{d (d^2-\Delta x^2) (6 \Delta x-5 R)}{5 \Delta x^3 R}\right)^2 \\ 
    \mathcal{G}_{cq}^2 &= \frac{ \left(\Delta x^2-4 d^2\right)^4 (6\Delta x-5 R_1)}{2560 d^2 \Delta x^3 R_1 R_2^3}\\
    \mathcal{G}_{cc}^2 &=  \frac{d^6}{4 R_1^3 R_2^3}
\end{align}
Thus, one can compute the values for known experiments given in Table~\ref{tab:two-body-benchmarks}.

\section{Bounding the Noise Kernels}
\label{app:experimental}
In an experiment, due to environmental interactions and a possibly fundamental noise, the two particles involved in a GME experiment would feel a random force
\begin{equation}
    \mathbf{E}[F^i_a(t) F^j_b(t')]= Q^{ij}_{ab} \delta(t-t') \ ,
\end{equation}
where
\begin{equation}
     Q^{ij}_{ab} = M_a M_b\int d^3 x d^3 y \  \rho_a(x-X_a) \rho_b(y-X_b) \partial_{x_i} \partial_{y_j}D(x-y) \ .
\end{equation}
Going to momentum space, we obtain
\begin{equation}
Q_{ab}=
M_aM_b\int \frac{d^3k}{(2\pi)^3}\,D(k)
k_z^2  \text{Re}\left[ \rho_a(k)\rho_b(k)^* e^{-ik\cdot(X_a-X_b)}\right].
\end{equation}

\subsubsection{Measured noise kernels and $\nu_{\text{exp}}$}
Compare this with
\begin{equation}
\Gamma_e=
4M_1M_2\int \frac{d^3k}{(2\pi)^3}\,D(k)
\sin\!\left(\frac{k_z\Delta x_1}{2}\right)
\sin\!\left(\frac{k_z\Delta x_2}{2}\right) \text{Re}\left[ \rho_1(k)\rho_2(k)^* e^{-ik\cdot(X_1-X_2)}\right] \ ,
\end{equation}
where we are leaving the superposition size arbitrary for the two bodies as well. For equal displacement ($\Delta x_1 = \Delta x_2 =\Delta x$, and we suppress the $zz$ indices for $Q$ from now on) 
\begin{equation}
    4\Gamma_l^{(aa)}2\Gamma_l^{(bb)}-\Gamma_e^2 \leq \Delta x^4 (Q_{aa}Q_{bb}-Q_{ab}^2) \ ,
\end{equation}
which follows from rewriting the decoherence matrices $\Gamma$ and the force covariance matrix $Q$ as
\begin{equation}
   \Delta x^2 Q- \Gamma = \Delta x^2\int \frac{d^3k}{(2\pi)^3} (1-\textrm{sinc}(k_z \Delta x/2)^2 ) D(k)   M(k) \geq 0 \  ,
\end{equation}
with $M(k)$ the appropriate matrix involving the mass densities. Hence,
\begin{equation}
    \nu_{\text{exp}}(\Delta x) \leq \frac{\Delta x^4 (Q_{11}Q_{22}-Q_{12}^2) }{\Gamma_{+,*}\Gamma_{-,*}} \ .
\end{equation}
For unequal superposition sizes, the cross-correlation inequality between $\Gamma_e$ and $Q_{ab}$ breaks down. Then, the only reliable upper bound comes from the worst-case scenario, i.e. $\Gamma_e=0$, meaning
    \begin{equation}
    \nu_{\text{exp}}(\Delta x_1,\Delta x_2) \leq \frac{\Delta x_1^2 \Delta x_2^2 (Q_{11}Q_{22})}{\Gamma_{+,*}\Gamma_{-,*}} \ .
\end{equation}
Conveniently, this simplifies greatly in the small $\Delta x_a$ limit, where, to leading order in the superposition sizes
\begin{equation}
    \Gamma_e \approx \Delta x_1 \Delta x_2 Q_{12} \ , \qquad 2\Gamma_l^{(a)} \approx \Delta x_a^2 Q_{aa}  \ ,
\end{equation}
yielding
\begin{equation}
    \lim_{\Delta x_a \to 0}\nu_{\text{exp}}= \frac{Q_{11}Q_{22}-Q_{12}^2}{Q_{11,*}Q_{22,*}-Q_{12,*}^2} \ ,
\end{equation}
where the starred quantities are those computed through the DP kernel. Experiments such as LISA Pathfinder often measure only the \emph{one-sided differential} acceleration power spectrum between the masses
\begin{equation}
    S^z_{\Delta a}(f>0) = \mathbf{E}[(a^z_1(f)-a^z_2(f))(a^z_1(f')-a^z_2(f'))] = 2 \left[\frac{Q_{11}}{M_1^2}+\frac{Q_{22}}{M_2^2}-2\frac{Q_{12}}{M_1 M_2}\right] \ ,
\end{equation}
which is frequency-independent since, we reiterate, we are working in the memoryless limit. Crucially, $S_{\Delta a}$ is insensitive to common-mode noise. Defining the ratio between correlated and uncorrelated noise to be
\begin{equation}
    r_{\parallel}\equiv \frac{Q_{12}}{\sqrt{Q_1 Q_2}} = \frac{S_{12}}{\sqrt{S_1 S_2}} \ ,
\end{equation}
the relative acceleration can be rewritten as
\begin{equation}
    S_{\Delta a}= S_1+S_2-2r\sqrt{S_1 S_2} \ ,
\end{equation}
meaning that unless one has an independent bound on the local acceleration, it is impossible to translate this into a bound on $\nu_{\text{exp}}$ without extra modelling assumptions. Indeed, we obtain
    \begin{equation}
    \nu_{\text{exp}}(\Delta x_1,\Delta x_2) \leq \frac{\Delta x_1^2 \Delta x_2^2 M_1^2 M_2^2 S_{\Delta a}^2}{\Gamma_{+,*}\Gamma_{-,*}(1-r_{\parallel})^2}.
    \label{eq:nu_adiff}
\end{equation}

\begin{figure*}[!t]
    \centering
    \includegraphics[width=0.45\linewidth]{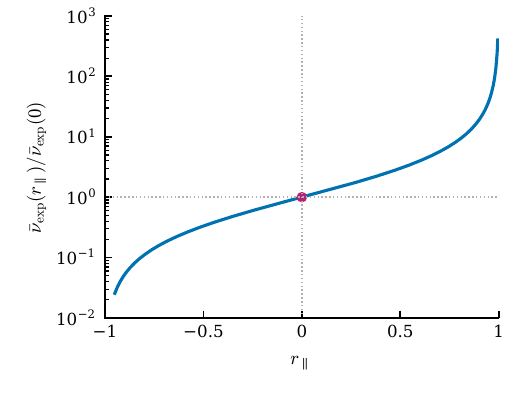}
    \caption{Dependence of the estimated $\nu_{\mathrm{exp}}$ on the assumed ratio $r_\parallel$ between correlated and uncorrelated noise. The solid circle represents the $r_\parallel=0$ choice made for all the figures in the main body.}
    \label{fig:r_par}
\end{figure*}

Without an independent bound on $r_\parallel$, $\nu_{\mathrm{exp}}$ cannot be computed reliably, as can be seen in Figure~\ref{fig:r_par}. However, it is physically sensible to assume that $|r_\parallel|\ll 1$, i.e. that any amount of nonlocal correlation is small. As we discuss in the next subsection, circumstantial evidence comes from studying the nonlocal DP kernel, for which, at the typical scales established by LISA Pathfinder, the correlation between the forces on the two masses is vanishingly small. Computing this for the three different types of experiments, we obtain
\begin{align}
    \nu_{qq} & \leq \frac{25 \Delta x^6 S_{\Delta a}^2}{4 \hbar^2 G^2 \left(\frac{(5 R-6 \Delta x)^2}{R^2}-\frac{25 \Delta x^6}{4 \left(d^3-d \Delta x^2\right)^2}\right)} \\
    \nu_{cq} & \leq \frac{5 \Delta x^3 S_{\Delta a}^2}{2 \hbar^2 G^2 \left(\frac{\frac{6 \Delta x}{R_1}-5}{R_2^3}-\frac{640 d^2 \Delta x^3}{\left(\Delta x^2-4 d^2\right)^4}\right)} \\
    \nu_{cc} &\leq
\frac{d^6 R_1^3 R_2^3 S_{\Delta a}^2}{16 \hbar^2 G^2 \left(d^6-4R_1^3 R_2^3\right)}
\end{align}
from which one can compute the values given in \textcolor{blue}{Table~\ref{tab:experimental-observables}}.

\subsubsection{Stochastic forces for different geometries}
Naturally, the stochastic force for each particle ($i=j$) reduces to the profile-averaged two-point function of the noise, which, after angular integration (recall that we assume spherically symmetric noise kernels), reads as a momentum-space integral:
\begin{equation}
     Q^{ij}_{a} \equiv Q^{ij}_{aa} = \frac{M_a^2\delta^{ij}}{6\pi^2}\int_0^\infty dk k^4D(k)|\rho_a(k)|^2 \  \ ,
\end{equation}
meaning that the noise is totally uncorrelated along orthogonal directions. Picking the minimal-noise DP kernel, and using isotropy ($Q^{xx}=Q^{yy}=Q^{zz}$), one then finds that the minimal noise for a single body is given by
\begin{equation}
    \sum_i Q_{a,*}^{ii} =  4\pi GM_i^2\int d^3 x \rho_i(x)^2 =  4\pi GM_i^2\int d^3 x \rho_i(x)^2\ , 
\end{equation}
which for a compact object of constant density reads
\begin{equation}
    Q_{a,*}^{ij} = \delta^{ij} \frac{4\pi GM_i^2}{3V} \ . 
\end{equation}

One can also explicitly compute the cross-correlation under the DP kernel for spheres of unequal masses and radii, since it amounts (thanks to their spherical symmetry) to the point-mass tidal tensor. Whilst some experiments (like LISA Pathfinder) do not use spheres, the approximation should be valid for distances $d>L$, where $L$ is the maximum length scale associated with the mass. At any rate, to compute the exact $Q_*$ for a specific geometry of the probe masses, one can readily use numerical integration. For non-overlapping spheres, we obtain
\begin{equation}
    Q^{ij}_{ab} = \frac{G M_i M_j}{d^3} (\delta^{ij}-3n^i n^j) \ ,
\end{equation}
meaning that the ratio between the correlated and uncorrelated noise along the centre-of-mass separation amounts to
\begin{equation}
    r_{\parallel,*} = \frac{Q^{zz}_{12}}{\sqrt{Q^{zz}_{1}Q^{zz}_{2}}} =-2 \frac{(R_1 R_2)^{3/2}}{d^3} \ ,
\end{equation}
which for identical spheres reduces to
\begin{equation}
     r_{\parallel,*} = -2 \left(\frac{R}{d}\right)^3 \ .
\end{equation}

\subsubsection{Decoherence experiments}

As noted in Ref.~\cite{minimal_fabiano_2026}, measuring $2\Gamma_l\leq\dot{\phi}_e$ in the GME experimental setup of choice is equivalent (and, therefore, similarly hard) to probing entanglement directly, since any correlated noise at fixed local rate \emph{helps} entanglement. However, the scenario considered by~\cite{minimal_fabiano_2026} is the worst case possible  this is the worst case scenario: we can have in fact access to $\Gamma_e$ itself.

Indeed, noise kernels are experimentally accessible by measuring the decoherence rates in different branches of quantum two-body experiments. Specifically
\begin{equation}
 |\rho_{LL,RR}(t)|\propto e^{-\Gamma_+t}, \qquad |\rho_{LR,RL}(t)|\propto e^{-\Gamma_-t} \ .
\end{equation}
However, quantum superposition experiments generically do not have good control over all the entries of the density matrix, and instead access only the local decoherence rates, which in turn affect the normalised coherence
\begin{equation}
    C_i(t)=
\frac{|\rho^{(i)}_{LR}(t)|}
{\sqrt{\rho^{(i)}_{LL}(t)\rho^{(i)}_{RR}(t)}}
\end{equation}
for a two-branch state. Interference visibility provides access to this quantity after accounting for branch populations and readout contrast. Plugging in our expression for the matrix elements of the reduced states, and ignoring the gravitational interaction
\begin{equation}
    C_i(t)=C_i(0)e^{-\Gamma_l^{(i)}t} \ ,
\end{equation}
with $C_0$ being the initial coherence that, for a conservative bound, can be fixed to unity. Let us denote by $U_i$ the estimated local decoherence coefficients extracted from the coherence times of massive superpositions.

We now consider a possible experiment with two copies of the same object, with the same noise control, and some geometric factor $\mathcal{G}_{qq}$. Then local upper bounds suffice to obtain (dropping the nonpositive term $-\Gamma_e^2$):
\begin{equation}
\nu_{\rm exp}\leq \bar\nu_{\rm exp}\equiv \frac{4U^2} {4\Gamma_{l,*}^2-\Gamma_{e,*}^2} =\frac{(U/\Gamma_{l,*})^2} {1-\mathcal G_{qq}^{-2}} \approx \frac{U^2}{\Gamma_{l,*}^2} , 
\end{equation}
where in the last step we used the fact that, as we have shown, the current experiments with the best geometric factors are limited by $\mathcal{G}\gg 1$.

In the absence of smearing, $\bar\nu_{\rm exp}<1$ would place the two-body noise below the reference DD trade-off floor. The stronger condition
\begin{equation}
\bar\nu_{\rm exp}<\frac{1}{\mathcal G_{qq}^2-1}
\end{equation}
would certify a noise level compatible with entanglement generation in the static Newtonian dephasing model with no smearing. Conveniently, and contrary to the LISA measurements, coherence experiments can be reliably translated into bounds on $\nu_{\mathrm{exp}}$ without needing to assume anything about the relative size of local and shared noise.

\subsection{Certifying the DD trade-off violation}
Although $\nu_{\mathrm{exp}}$ and the DD trade-off are closely related, there is no need to establish a reliable bound on $\nu_{\mathrm{exp}}$ to certify a violation of the DD trade-off. Crucially, certifying the violation can be done without having to assume anything about the ratio of local-to-nonlocal noise.

First, let us discuss the no-smearing case. For classical experiments, both relative-acceleration experiments like LISA Pathfinder and bounds on local acceleration can certify the violation. Indeed, under the DD trade-off
\begin{equation}
  S_{\Delta a} = 2\int d^3 k D(k) k_z^2 |\rho_1(k) e^{-ik\cdot X_1}-\rho_2(k)e^{-ik\cdot X_2}|^2 \geq  S_{\Delta a,*} \ ,
\end{equation}
i.e. the one computed with the DP kernel, since $D(k)\geq D_*(k)$. In short, no amount of common-mode noise can push the relative acceleration below the DP noise bound under the DD trade-off.

A similar computation can be carried out for local acceleration measurements, where $\sum_iQ_{a}^{ii} < \sum_iQ_{a, *}^{ii}$ implies that the DD trade-off is violated at least somewhere in the probed spectrum. In a similar spirit, if the local decoherence coefficients satisfy $U<\Gamma_{l,*}$, the DD trade-off cannot hold.

Introducing smearing is straightforward. Consider a generic positive observable $O$, measured in the lab, that can be expressed as
\begin{equation}
    O = \int d^3 k W_O(k) D(k) \ ,
\end{equation}
i.e. that samples the kernel in Fourier space, with $W_O(k)\geq 0$. For example
\begin{equation}
    W_{\Gamma_l}=
M^2|\rho( k)|^2
\left(1-\cos( k_z \Delta x)\right)
\end{equation}
for the local decoherence rates, or
\begin{equation}
    W_{S_{\Delta a}}= 8k_z ^2
|\rho(k)|^2
\sin^2\left(\frac{ k_z d}{2}\right),
\end{equation}
for the differential acceleration. As always in our notation, $O_{*}$ represents that observable computed with the DP kernel, whilst  $O_{g,*}$ the same DP observable but with smearing $g$:
\begin{equation}
    O_{g,*}= \int d^3 k W_O(k) D_*(k) |g(k)|^2\ .
\end{equation}
Then the violation of the DD trade-off through the observable $O$ can be checked. Then, by positivity and normalisation of $|g(k)|^2$, it follows that $O_{g,*}\leq O_{*} $. Calling $O_{\textrm{min}}$ the minimal value of $O$ attainable by minimising on the set of smearings that are not excluded experimentally, then defining
\begin{equation}
   \eta^T_O\equiv \frac{O_{\textrm{min}}}{O_{*}} \ , \eta_O\equiv \frac{O}{O_{*}} 
\end{equation}
implies that measuring
\begin{equation}
  \nu^{(O)}_{dd}\equiv \frac{\eta_O}{\eta^T_O} < \ 1,
\end{equation}
signifies a violation of the DD trade-off.

\subsubsection{Optimal smearings}
Although energy production is minimised by a parabolic smearing for a point-like mass~\cite{Piccione:2025zfb}, each observable (and choice of mass density) is minimised by a potentially different smearing function. The same parabolic function also minimises $\eta_{\Gamma_l}$ for a configuration with small $\Delta x$, but the minimisation procedure needs in principle to be performed for each observable of interest. We now follow the procedure outlined in Ref.~\cite{Piccione:2025zfb} to derive the optimal smearing for the quantities of interest, at least in certain limits where the computation can be done analytically. In particular, we assume point masses throughout, i.e. the smearing length is much greater than the size of the object. This is the only regime of interest, since in the opposite case smearing plays virtually no role, and the bare DP solution can be used instead as a good approximation of the threshold.

First, let the smearing function be isotropic and obey
\begin{equation}
    \int d^3 x \ g(x) =1 \ ,\qquad  g(x)\geq 0 \ ,\qquad \int d^3x x^2 g(x)= 3R_0^2 \ .
\end{equation}
Then, for an arbitrary observable $O$, the optimal fixed-variance kernel satisfies
\begin{equation}
    2\mathcal{F}^{-1}[W_O D_* g_{\mathrm{opt}}]+ \lambda + b\cdot x+\mu x^2 =\begin{cases}
        &=0, \ g_{\mathrm{opt}}>0 \ , \\
        & \geq 0 , \ g_{\mathrm{opt}}=0 \ ,
    \end{cases}
\end{equation}
with $\mathcal{F}^{-1}$ the inverse Fourier transform.

\begin{itemize}
    \item \textbf{Single-probe acceleration}: For the single-probe acceleration noise $S_{a}$, the problem is identical to the one solved in~\cite{Piccione:2025zfb}, yielding
\begin{equation}
    g^{a}_{\mathrm{opt}} = g^{\dot E}_{\mathrm{opt}}  =\frac{15}{8\pi (\sqrt{7}R_0)^5}\left[7R_0^2-r^2\right]_+ \ ,
\end{equation}
which is exact and requires no approximation.
    \item \textbf{Differential acceleration}: For the differential acceleration, a closed form is not always possible. For $d \gg R_0$, the contribution from the correlated acceleration under the DP kernel drops, meaning that the optimiser is once again the parabolic smearing. In the opposite limit, however,
    \begin{equation}
        S_{\Delta a,g } \approx \frac{2d^2}{5\kappa } \int d^3 x |\nabla g|^2 .
    \end{equation}
    Then, solving the stationary equation
    \begin{equation}
        -2\nabla^2 g +\lambda +\mu r^2 = 0
    \end{equation}
    and applying the normalisation conditions, we obtain a quartic polynomial in $r$. Summarising, in the two computable limits
    \begin{equation}
         g^{\Delta a}_{\mathrm{opt}}\approx
         \begin{cases}
              g^{a}_{\mathrm{opt}} \ , \qquad  &d\gg R_0 \\
              \frac{105}{32\pi (3R_0)^7}\left[9R_0^2-r^2 \right]^2_+ \qquad  &d\ll R_0
         \end{cases}
    \end{equation}
    \item \textbf{Local decoherence:} For the local decoherence, in the $\Delta x \ll R_0$ limit we have again the parabolic minimiser. In the opposite limit $\Delta x \gg R_0$, however, we obtain
    \begin{equation}
2U_g(r)+\lambda+\mu r^2=0
 \  ,
    \end{equation}
    with $U_g$ the gravitational self-energy of the smeared point-mass. Recognising $\nabla^2 U_g = -4\pi g$, it follows that $g$ must be a constant inside the support, yielding
    \begin{equation}
        g^{\Gamma}_{\mathrm{opt}} = 
        \begin{cases}
              g^{a}_{\mathrm{opt}} \ , \qquad  &\Delta x\ll R_0 \\
              \frac{3}{4\pi (\sqrt{5}R_0)^3}\Theta(\sqrt{5}R_0-r ) \qquad  &\Delta x\gg R_0
         \end{cases}
    \end{equation}
\end{itemize}
Since the smeared quantities are trivially monotonic in the smearing length $R_0$, it suffices to find the optimal smearing for each observable, and then take the largest allowed value for $R_0$ compatible with force experiments. 

\begin{figure}
    \centering
    \includegraphics[width=0.5\linewidth]{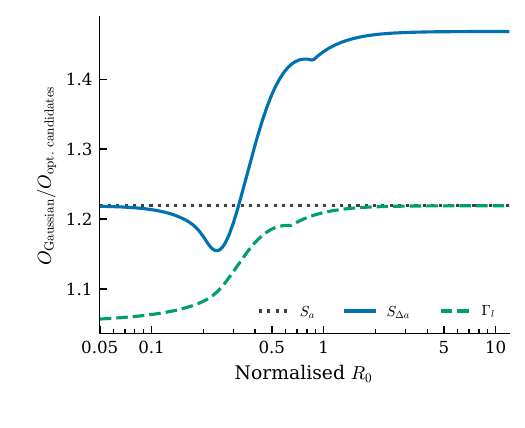}
    \caption{The ratio between the noise floors obtained with Gaussian and with optimal smearing for single-body acceleration ($S_a$), differential acceleration ($S_{\Delta a}$) and local decoherence rate ($\Gamma_l$). When only the asymptotic optimal smearings are available, the normalisation value is chosen at each point as the minimum of the two asymptotic choices. This shows that, although not strictly optimal, the Gaussian smearing can be used, for ease of computation, to yield back-of-the-envelope estimates for the observables of interest.}
    \label{fig:optimal_smearing}
\end{figure}

This avenue promises to find a relevant order-of-magnitude estimate of the lower bound. However, one would really need to do the constrained optimisation with a Lagrange multiplier for the force deviation itself at distance $d$, abandoning the ad hoc fixed-variance requirement. Still, a convenient choice for computations is the Gaussian smearing, for which we have a closed-form expression for all of these observables in the point-mass limit $R\ll R_0$:
\begin{align}
    S_{a,g_G} &= \frac{1}{12\kappa(\sqrt{\pi}R_0)^3} \, \\
     S_{\Delta a,g_G} &=   \frac{1}{4\kappa \pi R_0^3} \left[\frac{2}{3\sqrt{\pi}}+\frac{8 R_0^3\mathrm{erf}(d/2R_0)}{d^3}- \frac{e^{-d^2/(2R_0)^2}}{\sqrt{\pi}}\left( \frac{8 R_0^2}{d^2}+2\right) \right] \ , \\
     \Gamma_{l,g_G} &= \frac{M^2}{4\pi\kappa}\left[ \frac{1}{\sqrt{\pi}R_0} - \frac{\mathrm{erf}(\Delta x/2R_0)}{\Delta x}\right] \ .
\end{align}
As we show numerically in Figure~\ref{fig:optimal_smearing}, the Gaussian approximation yields sensible estimates for the observables of interest.

\subsection{Experimental status}
In this subsection, we discuss the various approximations we make to assign a set of proxy coefficients $(\nu,\mathcal{G},\eta_O)$ to a collection of experiments at the interface between quantum and gravitational physics. For some, extracting estimates of these coefficients is straightforward from experimental parameters and published results; for others, extra assumptions are necessary.
\subsubsection{LISA Pathfinder~\cite{Armano:2018kix}}
The LISA Pathfinder experiment is composed of two cubic gold–platinum test masses, with side length $L\sim 46\ \mathrm{mm}$, mass $M\sim 1.928 \ \mathrm{kg}$ and centre-of-mass separation $d\sim 376 \ \mathrm{mm}$. These data fix
\begin{equation}
    \mathcal{G} \approx 1.15 \times 10^3 \ ,
\end{equation}
meaning that the critical noise floor for entanglement (assessing the potential of the setup to entangle the masses in the small $\Delta x$ limit) is $\nu_{\textrm{crit}}\approx 7.59\times 10^{-7}$.

The setup uses interferometric displacement sensing to measure the differential acceleration spectrum along their separation axis. The latest, most sensitive measurements reached a noise floor of approximately $  \sqrt{S_{\Delta a}} \sim 1.74\ \mathrm{fm \ s^{-2}/\sqrt{Hz}}$~\cite{Armano:2018kix}. This provides a direct observable for comparison with the differential-noise DD trade-off floor. To simplify computations when estimating the DD trade-off floor, we model the test masses as spheres instead, fixing density and mass. Including the first-order corrections from smearing
\begin{equation}
S_{\Delta a}\approx \frac{4\hbar G}{R^3} \left[1-\frac{3R_0}{\sqrt\pi R} +\frac{2R_0^3}{\sqrt\pi R^3}\right] +\frac{8\hbar G}{d^3} \ ,
\end{equation}
which does not play a major role, since the upper bound on the nonlocal scale is necessarily one order of magnitude smaller than the size of the cubes, we find that the ratio between the measured and minimal differential-acceleration noise is
\begin{equation}
    \nu^{(\sqrt{S_{\Delta a}})}_{dd} =\frac{\sqrt{S_{\Delta a}}}{\sqrt{S_{\Delta a*}}}\approx 5.38\times 10^4
\end{equation}

However, inferring $\nu_{\rm exp}$ itself requires information about local noise or cross-correlations (i.e. $r_{\parallel}$). Having no independent bound on the common noise between the two masses, we need to make some approximations. In particular, as discussed in the previous section, we assume that such correlations are negligible, obtaining:
\begin{equation}
    \nu_{\mathrm{exp}}\lessapprox 6.20 \times 10^{18} \ .
\end{equation}
However, this bound is \emph{extremely} dependent on the $r_\parallel\ll 1$ assumption, with the upper bound diverging as $r_\parallel\to 1$, as can be seen in Equation~\ref{eq:nu_adiff}.

\subsubsection{Fuchs et al.~\cite{Fuchs:2023ajk}}
Fuchs et al. detect gravitational forces using a magnetically levitated composite particle of mass $M\approx 0.43 \mathrm{mg}$ (made up of three cubes of size $L\approx0.25 \ \mathrm{mm}$ and a sphere of similar radius) driven by external source masses -- which is found to be limited by a random acceleration noise of amplitude $\sqrt{S_{a}}\approx 1.16  \ \mathrm{nm \ s^{-2} \ Hz^{-1/2}}$.

For the DP reference, we approximate the composite particle by a homogeneous sphere of effective radius $R\approx 0.3 \ \mathrm{mm}$, yielding (ignoring smearing, which is negligible here)
\begin{equation}
    \nu^{(\sqrt{S_{ a}})}_{dd} \approx 3.76 \times 10^9 \ .
\end{equation}

From this experiment alone, neither $\mathcal{G}$ nor $\nu_{\mathrm{exp}}$ can be extracted directly, since their definitions strictly involve two objects. However, one could, if so desired, envision an experiment with a second copy of the particle placed at the optimal distance $d=2R$ for the plane-separated requirement, yielding
\begin{equation}
    \mathcal{G}^{(2)}=4 \ ,\qquad \nu_{\mathrm{crit}}^{(2)}=\frac{1}{15} \ ,
\end{equation}
where we used the index $(2)$ to indicate that these quantities refer to such an idealised experiment. Still, if each copy has local fundamental noise no larger than the measured bound, this yields
\begin{equation}
    \nu^{(2)}_{\mathrm{exp}}\leq  7.08\times 10^{30} \ .
\end{equation}

\subsubsection{Westphal et al.~\cite{measurement_westphal_2021}}
The Westphal et al. experiment measured the gravitational attraction of a small, independently controlled source using a miniature torsion pendulum. The setup involves three gold balls of approximately the same radius and mass: $M\approx 91 \ \mathrm{mg}$ and $R \approx 1.1\,\mathrm{mm}$; one acting as source, another as the test mass, and the final one as the counterbalance in the torsion pendulum. The source–test centre separation varies $2.5 \ \mathrm{mm} \lessapprox d \lessapprox 5.8 \ \mathrm{mm}$, with source modulation at $12.7\,\mathrm{mHz}$. 

This experimental setup does not fit neatly into our construction. However, a proxy for the geometric factor of the experiment comes from considering only the gravitationally relevant particles (test and source). Taking the equilibrium position of the torsion pendulum as a proxy gives
\begin{equation}
    \mathcal{G} \approx 32 \ .
\end{equation}

The experiment is sensitive to an acceleration modulation of approximately $\sigma_a\approx 3\times10^{-12} \ \mathrm{m\,s^{-2}}$. Translating the torsion noise into a bound on the local acceleration of the masses requires a careful analysis, but we attempt here to derive a meaningful estimate for the noise bound that can be extracted from the observations. The measured torsional spectrum is mainly probing the relative acceleration between test and counterbalance masses
\begin{equation}
    \tau = M L (a_t-a_c)
\end{equation}
where $L$ is the length of each lever arm of the torsion pendulum, i.e. $L\approx 20 \ \mathrm{mm}$ -- although in principle the noise in the position of the force can also contribute. The quoted uncertainty in the torque after $T=350 \ \mathrm{hr}$ of averaging can then be used to extract a proxy for
\begin{equation}
    \nu^{(\sqrt{S_{\Delta a}})}=\sqrt{\frac{\sigma_a^2 T}{S_{\Delta a *}}}\approx 7.9 \times10^9 \ ,
\end{equation}
where we still used the unsmeared DP kernel, for reference, since the size of the spheres is larger than the maximum allowed $R_0$ from gravitational measurements.

As before, extracting $\nu_{\mathrm{exp}}$ from a differential acceleration measurement requires extra assumptions on the correlation between the local acceleration noise of the masses. We assume that the source mass could attain the same local acceleration-noise level as the test and counterbalance masses and, again, adopt the sensible assumption $r_\parallel \ll 1$:
\begin{equation}
    \nu_{\mathrm{exp}}\lessapprox 2.1 \times 10^{35}
\end{equation}

\subsubsection{Panda et al.~\cite{Panda:2022gtw}}
The Panda et al. experiment studied the coherence time of individual $^{133}\mathrm{Cs}$ atoms in an optical lattice. The atoms were split into a superposition with branch separation $\Delta x \approx 1.9 \ \mathrm{\mu}m$ ($\Delta x \ll R_0$ for our worst-case scenario); the observable contrast was measured to be $C(70s)\approx 0.03$.

A rough bound on the local decoherence can be readily extracted from the loss of contrast
\begin{equation}
    \nu^{(\Gamma_l)}=\frac{\eta^{\Gamma_l}}{\eta^{\Gamma_l}_T}\approx 1.32 \times 10^{29} \ ,
\end{equation}
where here, since the smearing scale becomes relevant, we have used the Gaussian-smeared result with $R_0 \approx 2.4 \ \mathrm{m m}$ instead.

Again, one could consider an experiment with a second copy of the atom placed at the optimal distance $d=2R+\Delta x$ (taking, for the sake of this (in any case idealised) discussion, $R$ to be the atomic radius). Then, for these parameters we obtain
\begin{equation}
    \mathcal{G}^{(2)} \approx 4.8 \ ,
\end{equation}
and, with no further assumption, the associated
\begin{equation}
    \nu_{\textrm{exp}}^{(2)}\lessapprox1.15 \times10^{29} \ , \qquad \nu_{\mathrm{crit}}^{(2)}\approx0.05 \ .
\end{equation}
Even in this idealised case, in the absence of smearing, the noise needs to be reduced by several orders of magnitude to have a setup with a chance to detect GME.

\subsubsection{Kovachy et al.~\cite{quantum_kovachy_2015}}
Atomic fountain experiments can also potentially provide interesting bounds. The Kovachy et al. experiment involves ultracold $^{87}\mathrm{Rb}$ atoms which, through a sequence of laser pulses, are first put in a superposition of branches separated by a maximum of $\Delta x \approx 54 \ \mathrm{cm}$ ($\Delta x \gg R_0$ for our worst-case scenario), and then recombined after $T\approx 2.08\ \mathrm{s}$.

The reported contrast at recombination, $C\approx0.28$, yields, in the worst-case scenario,
\begin{equation}
     \nu^{(\Gamma_l)}\approx 1.99 \times 10^{23} \ .
\end{equation}
Again, we can imagine an idealised experimental setup, where two fountains are set up side by side at the minimal separation, yielding once more $\mathcal{G}^{(2)}\approx 4.8$. In turn, this implies
\begin{equation}
    \nu_{\mathrm{exp}}^{(2)}\lessapprox 1.4 \times 10^{32}
\end{equation}

\subsubsection{Bose et al. proposal~\cite{bose_spin_2017}}
Let us consider the first practical experimental proposal for the BMV setup, in which two microspheres of radius $R\sim 1\mu\text{m}$ and mass $m\sim 10^{-14}\text{kg}$ are imagined in a superposition of width $\Delta x\sim 250\mu\text{m}$ at a distance of $d\sim450\mu\text{m}$~\cite{bose_spin_2017}. This configuration corresponds to a geometric factor
\begin{equation}
    \mathcal{G}_{\text{BMV}}\simeq 1.2 \times 10^3 \ ,
\end{equation}
which is significant. 

With this setup, the critical noise threshold lies well below the DD trade-off line in the absence of smearing, at $\nu_{\text{crit},BMV}=1/(\mathcal{G}^2_{\text{BMV}}-1)\simeq 6.9\times10^{-7}$. Note that this is in fact smaller than the best-case scenario proposed in the paper, where the source of decoherence, assumed to be uncorrelated ($\Gamma_e=0$) collisions of the masses with gas particles at pressure $P=10^{-15}\text{Pa}$ and temperature $T=0.15\text{K}$, has a collisional decoherence time of the order of $3.5$~s~\cite{bose_spin_2017}. Compared with $\Gamma_{l,*}\approx76~\mathrm{s^{-1}}$, this leads to $\nu_{\text{BMV,env}}=(\Gamma_l/\Gamma_{l,*})^2\sim10^{-5}$ instead.
\begin{figure*}[!t]
    \centering
    \includegraphics[width=0.95\linewidth]{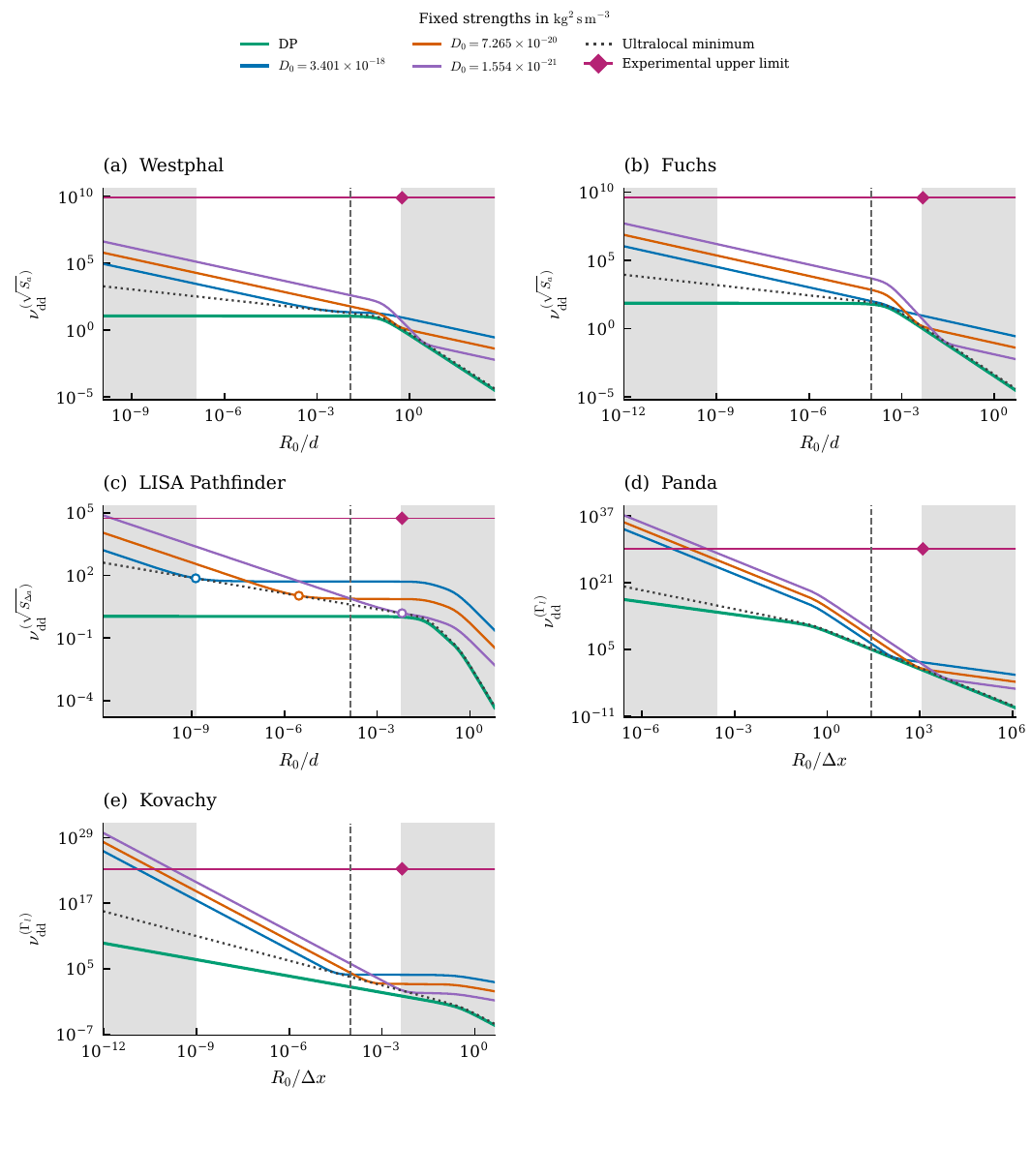}
    \caption{Minimal theoretical prediction, under the DD trade-off, of the relevant observable measured by the experiments of interest, as a function of the Gaussian smearing scale. The green line is the minimal noise DP model, whilst the coloured lines represent the fully local CQ model, which enjoys a further free parameter $D_l$. The black dashed represents the upper bound on the Bessel smearing function. \label{fig:bounds_app}}
   
\end{figure*}

\subsubsection{Other relevant experiments}

Our analysis aimed to provide a simple recipe for assessing how far a given setup is from being a potential candidate for observing GME. Figure~\ref{fig:bounds_app} explores the bounds on the DD trade-off for the set of experiments we have discussed so far.

However, several relevant experiments cannot be assigned the parameters introduced above without additional modelling. In some cases, the difficulty lies in the geometry or the measured mechanical mode; in others, the reported observable does not directly constrain the decoherence or diffusion coefficients required by our comparison.
\begin{itemize}
    \item \textbf{Lee et al.~\cite{new_lee_2020}}: a test of the gravitational inverse-square law using a torsion balance and a rotating, patterned attractor. Their measured torque harmonics depend on the full mass distributions and cannot be faithfully represented by a pair of homogeneous spheres. Moreover, agreement with the Newtonian mean torque does not itself bound stochastic torque fluctuations. These measurements are particularly relevant for restricting modifications of the interaction due to smearing, although translating their published Yukawa limits into a constraint on a Gaussian smearing length requires a separate analysis. Our force-deviation bound, moreover, strictly applies to spherical objects, and would require a generalisation.
    \item \textbf{Overstreet et al.~\cite{observation_overstreet_2022}}: measurement of the gravitational Aharonov--Bohm effect. The differential phase uncertainty constrains a combination of phase fluctuations and their correlations, rather than an individual local coefficient, making any comparison extremely dependent on extra assumptions. 
    \item \textbf{Pedalino et al.~\cite{probing_pedalino_2026}}: this experiment demonstrates centre-of-mass interference of massive sodium clusters in a Talbot--Lau interferometer. Although the measured visibility constrains modifications that suppress matter-wave interference, the experiment involves multiple diffraction paths rather than two localized spatial branches -- so it lies outside our two-branch approximation.
    \item \textbf{Vinante et al.~\cite{Vinante:2020gow}}: an experiment bounding additional force noise through an ultracold cantilever. Their measured fluctuations can be compared with a model-specific diffusion prediction, but the relevant observable is the force projected onto a flexural mode of a composite structure, which requires additional modelling to be translated into our formalism.
    \item \textbf{XENONnT collaboration~\cite{XENON:2025iov}}: constrains collapse-induced radiation rather than noise directly. Translating its photon spectrum into the coefficients considered here requires a microscopic model of the coupling to charged constituents, together with assumptions about the temporal spectrum of the noise. Although completely different in nature from GME experiments, such experiments can provide formidable bounds on the DD trade-off.
\end{itemize}

More generally, cantilever and other mechanical-resonator measurements constrain noise in a selected mode rather than the translational diffusion of an isolated rigid body. Applying our framework requires weighting the spatial noise kernel by the mode's displacement field and the device's mass distribution. Such experiments remain valuable tests of additional diffusion, but reducing them to the two-body parameters used here would introduce assumptions that warrant a dedicated analysis.

\begin{table*}[t]
\centering
\small
\setlength{\tabcolsep}{4pt}
\renewcommand{\arraystretch}{1.2}
\begin{tabular}{lccccccc}
\hline\hline
Experiment & $\Delta x_1$ & $\Delta x_2$ & $d$ & $R_1$ & $R_2$ & $M_1$ & $M_2$ \\
\hline
LISA Pathfinder & $0$ & $0$ & $376\,\mathrm{mm}$ & $28.5\,\mathrm{mm}$ & $28.5\,\mathrm{mm}$ & $1.928\,\mathrm{kg}$ & $1.928\,\mathrm{kg}$ \\
Fuchs et al. & $0$ & $0$ & --- & $0.30\,\mathrm{mm}$ & --- & $0.43\,\mathrm{mg}$ & $3\times2.45\,\mathrm{kg}$ \\
Westphal et al. & $0$ & $0$ & $2.5$--$5.8\,\mathrm{mm}$ & $1.04\,\mathrm{mm}$ & $1.07\,\mathrm{mm}$ & $90.7\,\mathrm{mg}$ & $92.1\,\mathrm{mg}$ \\
Panda et al. & $1.9\,\mu\mathrm{m}$ & --- & --- & --- & --- & $133\,\mathrm{a.m.u.}$ & --- \\
Kovachy et al. & $0.54\,\mathrm{m}$ & --- & --- & --- & --- & $87\,\mathrm{a.m.u.}$ & --- \\
\hline\hline
\end{tabular}
\caption{Parameters of experiments of interest, with spherical approximations for the size of the test masses. Body 1 is the sensor and body 2 the source, where applicable. A zero branch separation denotes the absence of a prepared spatial superposition. Dashes denote unassigned parameters. \label{tab:experimental-parameters}}
\end{table*}

\begin{table*}[t]
\centering
\small
\setlength{\tabcolsep}{4pt}
\renewcommand{\arraystretch}{1.2}
\begin{tabular}{lllc}
\hline\hline
Experiment & Observable & Experimental input & $\nu_{\mathrm{dd}}^{(O)}$ \\
\hline
LISA Pathfinder & \parbox[t]{0.20\textwidth}{\raggedright Differential acceleration amplitude spectrum} & \parbox[t]{0.38\textwidth}{\raggedright $\sqrt{S_{\Delta a}}\simeq1.74\,\mathrm{fm\,s^{-2}\,Hz^{-1/2}}$} & $5.38\times10^{4}$ \\
Fuchs et al. & \parbox[t]{0.20\textwidth}{\raggedright Local acceleration amplitude spectrum} & \parbox[t]{0.38\textwidth}{\raggedright $\sqrt{S_a}\simeq1.16\,\mathrm{nm\,s^{-2}\,Hz^{-1/2}}$} & $3.76\times10^{9}$ \\
Westphal et al. & \parbox[t]{0.20\textwidth}{\raggedright Statistical uncertainty in fitted gravitational signal} & \parbox[t]{0.38\textwidth}{\raggedright $\sigma_a\simeq 3\times10^{-12}\,\mathrm{m\,s^{-2}}$, $T_{\mathrm{int}}=350\,\mathrm{h}$} & $7.9\times 10^9$ \\
Panda et al. & \parbox[t]{0.20\textwidth}{\raggedright Interference contrast} & \parbox[t]{0.38\textwidth}{\raggedright $C(70\,\mathrm{s})\simeq0.03$, $U\simeq0.050\,\mathrm{s^{-1}}$} & $1.38\times10^{29}$ \\
Kovachy et al. & \parbox[t]{0.20\textwidth}{\raggedright Interference contrast} & \parbox[t]{0.38\textwidth}{\raggedright $C(2.08\,\mathrm{s})\simeq0.28$, $U\simeq0.61\,\mathrm{s^{-1}}$} & $1.99\times10^{23}$ \\
\hline\hline
\end{tabular}
\caption{Experimental observables and approximate ratios to the decoherence--diffusion reference. The atomic reference estimates use Gaussian smearing with $R_0=10\,\mu\mathrm{m}$; for the larger objects, smearing is a subleading effect. The measurement mappings and their assumptions are discussed in Appendix~\ref{app:experimental}.\label{tab:experimental-observables}}
\end{table*}

\begin{table*}[t]
\centering
\small
\setlength{\tabcolsep}{6pt}
\renewcommand{\arraystretch}{1.2}
\begin{tabular}{lcccc}
\hline\hline
Experiment or proposal & $\mathcal{G}$ & $\nu_{\mathrm{crit}}$ & $\bar{\nu}_{\mathrm{exp}}$ & $\bar{\chi}_{\mathrm{GME}}$ \\
\hline
LISA Pathfinder & $1.15\times10^{3}$ & $7.6\times10^{-7}$ & $6.2\times10^{18}$ & $8.2\times10^{24}$ \\
Fuchs et al.$^{\ast}$ & $4$ & $6.7\times10^{-2}$ & $7.1\times10^{30}$ & $1.1\times10^{32}$ \\
Westphal et al. & $32$ & $9.8\times10^{-4}$ & $2.1\times 10^{35}$ & $2.1\times 10^{38}$ \\
Panda et al.$^{\ast}$ & $4.8$ & $4.5\times10^{-2}$ & $1.2\times10^{29}$ & $2.5\times10^{30}$ \\
Kovachy et al.$^{\ast}$ & $4.8$ & $4.5\times10^{-2}$ & $1.4\times10^{32}$ & $3.1\times10^{33}$ \\
Bose et al. proposal & $1.2\times10^{3}$ & $6.9\times10^{-7}$ & $\nu_{\mathrm{env}}\simeq1.4\times10^{-5}$ & $\chi_{\mathrm{env}}\simeq20$ \\
\hline\hline
\end{tabular}
\caption{Conditional two-body benchmarks towards GME in the absence of smearing -- and hence assuming no deviation from the inverse-square law is observed at the relevant scale. For experimental bounds, $\nu_{\mathrm{exp}}\leq\bar{\nu}$ and $\bar{\chi}=\bar{\nu}/\nu_{\mathrm{crit}}$. An asterisk denotes a hypothetical static two-copy configuration; see the discussion above. \label{tab:two-body-benchmarks}}
\end{table*}

\twocolumngrid

\bibliography{apssamp}

\end{document}